\documentclass[ALICE,manyauthors]{cernphprep}
\usepackage[comma,square,numbers,sort&compress]{natbib}
\usepackage{hyperref}
\usepackage{lineno}
\usepackage{xspace}
\usepackage{xcolor}
\usepackage{xcolor} 
\usepackage{soul} 
\usepackage{mathtools}
\usepackage{multirow}
\usepackage{amsmath}
\usepackage{graphicx}
\usepackage[utf8]{inputenc}
\usepackage[T1]{fontenc}

\begin{document}
%

\newcommand{\pp}           {pp\xspace}
\newcommand{\ppbar}        {\mbox{$\mathrm {p\overline{p}}$}\xspace}
\newcommand{\XeXe}         {\mbox{Xe--Xe}\xspace}
\newcommand{\PbPb}         {\mbox{Pb--Pb}\xspace}
\newcommand{\pA}           {\mbox{pA}\xspace}
\newcommand{\pPb}          {\mbox{p--Pb}\xspace}
\newcommand{\AuAu}         {\mbox{Au--Au}\xspace}
\newcommand{\dAu}          {\mbox{d--Au}\xspace}

\newcommand{\s}            {\ensuremath{\sqrt{s}}\xspace}
\newcommand{\snn}          {\ensuremath{\sqrt{s_{\mathrm{NN}}}}\xspace}
\newcommand{\pt}           {\ensuremath{p_{\rm T}}\xspace}
\newcommand{\meanpt}       {$\langle p_{\mathrm{T}}\rangle$\xspace}
\newcommand{\ycms}         {\ensuremath{y_{\rm CMS}}\xspace}
\newcommand{\ylab}         {\ensuremath{y_{\rm lab}}\xspace}
\newcommand{\etarange}[1]  {\mbox{$\left | \eta \right |~<~#1$}}
\newcommand{\yrange}[1]    {\mbox{$\left | y \right |~<~#1$}}
\newcommand{\dndy}         {\ensuremath{\mathrm{d}N_\mathrm{ch}/\mathrm{d}y}\xspace}
\newcommand{\dndeta}       {\ensuremath{\mathrm{d}N_\mathrm{ch}/\mathrm{d}\eta}\xspace}
\newcommand{\avdndeta}     {\ensuremath{\langle\dndeta\rangle}\xspace}
\newcommand{\dNdy}         {\ensuremath{\mathrm{d}N_\mathrm{ch}/\mathrm{d}y}\xspace}
\newcommand{\Npart}        {\ensuremath{N_\mathrm{part}}\xspace}
\newcommand{\Ncoll}        {\ensuremath{N_\mathrm{coll}}\xspace}
\newcommand{\dEdx}         {\ensuremath{\textrm{d}E/\textrm{d}x}\xspace}
\newcommand{\RpPb}         {\ensuremath{R_{\rm pPb}}\xspace}

\newcommand{\nineH}        {$\sqrt{s}~=~0.9$~Te\kern-.1emV\xspace}
\newcommand{\seven}        {$\sqrt{s}~=~7$~Te\kern-.1emV\xspace}
\newcommand{\twoH}         {$\sqrt{s}~=~0.2$~Te\kern-.1emV\xspace}
\newcommand{\twosevensix}  {$\sqrt{s}~=~2.76$~Te\kern-.1emV\xspace}
\newcommand{\five}         {$\sqrt{s}~=~5.02$~Te\kern-.1emV\xspace}
\newcommand{\twosevensixnn}{$\sqrt{s_{\mathrm{NN}}}~=~2.76$~Te\kern-.1emV\xspace}
\newcommand{\fivenn}       {$\sqrt{s_{\mathrm{NN}}}~=~5.02$~Te\kern-.1emV\xspace}
\newcommand{\LT}           {L{\'e}vy-Tsallis\xspace}
\newcommand{\GeVc}         {Ge\kern-.1emV/$c$\xspace}
\newcommand{\MeVc}         {Me\kern-.1emV/$c$\xspace}
\newcommand{\TeV}          {Te\kern-.1emV\xspace}
\newcommand{\GeV}          {Ge\kern-.1emV\xspace}
\newcommand{\MeV}          {Me\kern-.1emV\xspace}
\newcommand{\GeVmass}      {Ge\kern-.2emV/$c^2$\xspace}
\newcommand{\MeVmass}      {Me\kern-.2emV/$c^2$\xspace}
\newcommand{\lumi}         {\ensuremath{\mathcal{L}}\xspace}

\newcommand{\ITS}          {\rm{ITS}\xspace}
\newcommand{\TOF}          {\rm{TOF}\xspace}
\newcommand{\ZDC}          {\rm{ZDC}\xspace}
\newcommand{\ZDCs}         {\rm{ZDCs}\xspace}
\newcommand{\ZNA}          {\rm{ZNA}\xspace}
\newcommand{\ZNC}          {\rm{ZNC}\xspace}
\newcommand{\SPD}          {\rm{SPD}\xspace}
\newcommand{\SDD}          {\rm{SDD}\xspace}
\newcommand{\SSD}          {\rm{SSD}\xspace}
\newcommand{\TPC}          {\rm{TPC}\xspace}
\newcommand{\TRD}          {\rm{TRD}\xspace}
\newcommand{\VZERO}        {\rm{V0}\xspace}
\newcommand{\VZEROA}       {\rm{V0A}\xspace}
\newcommand{\VZEROC}       {\rm{V0C}\xspace}
\newcommand{\Vdecay} 	   {\ensuremath{V^{0}}\xspace}

\newcommand{\ee}           {\ensuremath{e^{+}e^{-}}} 
\newcommand{\pip}          {\ensuremath{\pi^{+}}\xspace}
\newcommand{\pim}          {\ensuremath{\pi^{-}}\xspace}
\newcommand{\kap}          {\ensuremath{\rm{K}^{+}}\xspace}
\newcommand{\kam}          {\ensuremath{\rm{K}^{-}}\xspace}
\newcommand{\pbar}         {\ensuremath{\rm\overline{p}}\xspace}
\newcommand{\kzero}        {\ensuremath{{\rm K}^{0}_{\rm{S}}}\xspace}
\newcommand{\lmb}          {\ensuremath{\Lambda}\xspace}
\newcommand{\almb}         {\ensuremath{\overline{\Lambda}}\xspace}
\newcommand{\Om}           {\ensuremath{\Omega^-}\xspace}
\newcommand{\Mo}           {\ensuremath{\overline{\Omega}^+}\xspace}
\newcommand{\X}            {\ensuremath{\Xi^-}\xspace}
\newcommand{\Ix}           {\ensuremath{\overline{\Xi}^+}\xspace}
\newcommand{\Xis}          {\ensuremath{\Xi^{\pm}}\xspace}
\newcommand{\Oms}          {\ensuremath{\Omega^{\pm}}\xspace}
\newcommand{\degree}       {\ensuremath{^{\rm o}}\xspace}

\begin{titlepage}
\PHyear{2020}       
\PHnumber{098}      
\PHdate{29 May}  

\title{Constraining the Chiral Magnetic Effect with charge-dependent
  azimuthal correlations in Pb--Pb collisions at
  $\sqrt{\it{s}_{\mathrm{NN}}}$ = 2.76 and 5.02~TeV} 
\ShortTitle{Chiral Magnetic Effect constrains at the LHC}   

\Collaboration{ALICE Collaboration\thanks{See Appendix~\ref{app:collab} for the list of collaboration members}}
\ShortAuthor{ALICE Collaboration} 

\begin{abstract}
Systematic studies of charge-dependent two- and three-particle
correlations in Pb--Pb collisions at $\sqrt{\it{s}_\mathrm{{NN}}} = $
2.76~and 5.02~TeV used to probe the Chiral Magnetic Effect (CME) are
presented. These measurements are performed for charged particles in
the pseudorapidity ($\eta$) and transverse momentum ($p_{\rm{T}}$)
ranges $\left|\eta \right| < 0.8$ and $0.2 < p_{\mathrm{T}} < 5$~GeV/$c$.
A significant charge-dependent signal that becomes more pronounced for
peripheral collisions is reported for the CME-sensitive correlators
$\gamma_{1,1} = \langle \cos (\varphi_{\alpha} +\varphi_{\beta} - 2\Psi_{2}) \rangle$
and $\gamma_{1,-3} = \langle \cos (\varphi_{\alpha} -3\varphi_{\beta} + 2\Psi_{2}) \rangle$. 
The results are used to estimate the contribution of background
effects, associated with local charge conservation coupled to
anisotropic flow modulations, to measurements of the CME. A blast-wave
parametrisation that incorporates local charge conservation 
tuned to reproduce the centrality dependent background effects is not
able to fully describe the measured $\gamma_{1,1}$. Finally, the
charge and centrality dependence of mixed-harmonics three-particle
correlations, of the form  
$\gamma_{1,2} = \langle \cos (\varphi_{\alpha} +2\varphi_{\beta} - 3\Psi_{3}) \rangle$, 
which are insensitive to the CME signal, verify again that background
contributions dominate the measurement of $\gamma_{1,1}$.

\end{abstract}

\end{titlepage}

\setcounter{page}{2} 

\section{Introduction} 
\label{Section:Introduction}
Heavy-ion collisions at ultra-relativistic energies are used to study
the phase transition from a deconfined Quark--Gluon Plasma (QGP) 
state~\cite{Shuryak:1984nq,Cleymans:1985wb,Bass:1998vz}  
to ordinary nuclear matter. The transition is expected to occur at high 
values of temperature and energy density, which is also supported by 
quantum chromodynamics (QCD) calculations on the 
lattice~\cite{Borsanyi:2010cj,Bhattacharya:2014ara}. The main aim of the 
heavy-ion program at the Large Hadron Collider (LHC) is to study the QGP 
properties, such as the equation of state, the speed of sound in the medium 
and the value of the ratio of shear viscosity to entropy density ($\eta/s$).

It was soon realised that heavy-ion collisions also allow for
studies of novel QCD phenomena associated with parity (P) violation
effects in strong interactions~\cite{Lee:1973iz,Lee:1974ma}. These
effects are catalysed by the presence of a strong magnetic field that
develops in the early stages of heavy-ion collisions. This field is
created by the motion of the charged nucleons of the incoming ions in
a non-central collision, i.e. a collision with a large impact
parameter, defined as the distance between the centers of the two
colliding nuclei in the transverse plane. The magnitude of this field
can reach values of  $10^{18}$~Gauss~\cite{Bzdak:2011yy}, making it
the strongest magnetic field created by any experiment on earth. The
direction of the magnetic field is along the system's angular momentum
and perpendicular to the reaction plane. The latter is the plane
defined by the impact parameter  vector and the beam direction.

The potential to observe parity violation in the strong interaction using 
ultra-relativistic heavy-ion collisions has first been discussed 
in Refs.~\cite{Morley:1983wr,Kharzeev:1998kz,Kharzeev:1999cz} and was further 
reviewed in Ref.~\cite{Kharzeev:2013ffa,Kharzeev:2015kna}. In QCD,
this symmetry violation originates from the possibility that the QGP can carry net 
chirality~\cite{Kharzeev:2007tn,Fukushima:2008xe,Kharzeev:2007jp}, 
characterised by a non-zero value of the axial chemical potential 
$\mu_5$, i.e. reflecting the imbalance between left-- and 
right--handed fermions in the system. Depending on the sign of 
$\mu_5$ the QGP will have an excess of either left-- ($\mu_5 < 0)$ 
or right--handed ($\mu_5 > 0$) (anti-)quarks. In the presence of the 
strong magnetic field, the spins of (anti-)quarks tend to align along the 
direction of the field, creating a spin polarisation effect. This in turn leads to 
the development of a vector current along the direction of the magnetic 
field and the creation of an electric dipole moment of QCD matter. The 
experimental search for these effects has intensified lately, following the 
realisation that the subsequent creation of charged hadrons results in an 
experimentally accessible charge separation along the direction 
of this magnetic field, and perpendicular to the reaction plane. This 
phenomenon is called the Chiral Magnetic Effect (CME) and its existence was 
recently reported in semimetals like zirconium pentatelluride 
($\rm ZrTe_5$)~\cite{Li:2014bha}.

The resulting charge separation can be identified by studying the P-odd sine 
terms in the Fourier decomposition of the particle azimuthal 
distribution~\cite{Voloshin:1994mz} according to

\begin{eqnarray}
\frac{dN}{d\varphi_{\alpha}} \sim 1 + 2\sum_n \left[v_{\it n,\alpha}
  \cos(n \Delta \varphi_{\alpha}) + \mathit a_{\it n,\alpha} \sin(n \Delta 
\varphi_{\alpha})\right],
\label{Eq:Fourier}
\end{eqnarray}

\noindent where $\Delta \varphi_{\alpha}=\varphi_\alpha - \Psi_{\rm RP}$ is the 
azimuthal angle $\varphi_\alpha$ of the particle of type $\alpha$
(either positively or negatively charged particles) relative to 
the reaction plane angle $\Psi_{\rm RP}$. The coefficient
$v_{n,\alpha}$ is the $n$-th order Fourier harmonic, averaged over all
events, and characterises the anisotropies in momentum space. The reaction
plane is not an experimental observable but can be approximated 
by the second-order symmetry plane, $\Psi_2$, determined by the
direction of  the beam and the axis of the maximal particle density in
the elliptic azimuthal anisotropy. This symmetry plane and more
generally the plane angles of different order $\Psi_n$, estimated in
each event, are introduced to account for the event-by-event
fluctuations in the initial energy density of a heavy-ion 
collision~\cite{Manly:2005zy,Bhalerao:2006tp,Alver:2008zza,Alver:2010gr,Alver:2010dn}. 
In case of a smooth distribution of matter produced in the overlap zone, the angle 
$\Psi_2$ coincides with that of the reaction plane, i.e. $\Psi_2 = \Psi_{\mathrm{RP}}$. 
The leading order P-odd coefficient $a_{1,\alpha}$ reflects the magnitude of the 
effects from local parity violation, while higher 
orders ($a_{\it n,\alpha}$ for $n > 1$) describe the specific shape in
azimuth. However, the chiral imbalance that leads to the  
creation of the CME changes from event to event and the event average 
$\langle a_{1,\alpha} \rangle$ will be consistent with zero. 
Consequently, the effect can be detected only by correlation studies.

In Ref.~\cite{Voloshin:2004vk}, it was suggested that a suitable way
to probe the CME is via a two-particle correlation technique relative
to the second-order symmetry plane of the form $\langle
\cos(\varphi_{\alpha} + \varphi_{\beta} - 2\Psi_{2}) \rangle$, where  
the brackets indicate an average over all events. Here, $\alpha$ and
$\beta$ denote particles with the same or opposite electric
charge. The advantage of using this expression is that  
it probes correlations between two leading order P-odd coefficients
$a_{1,\alpha}$ and $a_{1,\beta}$ which do not trivially average to 0
over all events (see Section~\ref{Section:Analysis} for the
discussion). In addition, the observable is constructed as the
difference between correlations in- and out-of plane   
which is expected to significantly suppress parity-conserving
background effects. In order to independently evaluate the
contributions from correlations in- and out-of plane one  
measures at the same time a two-particle correlator of the form 
$\langle \cos(\varphi_{\alpha} - \varphi_{\beta}) \rangle$. Section~\ref{Section:Analysis} 
contains a detailed discussion about all these correlators.

Experimental results for charged particles in both Pb--Pb collisions
at $\sqrt{s_{\rm NN}} = 2.76$~TeV at the LHC~\cite{Abelev:2012pa} and
in Au--Au collisions up to $\sqrt{s_{\rm NN}} = 0.2$~TeV at the 
Relativistic Heavy-Ion 
Collider (RHIC)~\cite{Abelev:2009ac,Abelev:2009ad,Adamczyk:2013kcb,Adamczyk:2013hsi,Adamczyk:2014mzf} 
are consistent with the expectation for a charge separation relative
to the reaction plane due to the existence of parity violating
effects. However, these measurements could be dominated by background
effects whose sources have not been fully quantified yet. One of the
first attempts to provide a quantitative estimate of the background in
the measurement of the CME sensitive correlator  (i.e. $\langle
\cos(\varphi_{\alpha} + \varphi_{\beta} - 2\Psi_{2}) \rangle$)
identified the sources as originating from local charge conservation
coupled to the elliptic flow modulation quantified by
$v_2$~\cite{Schlichting:2010qia,Pratt:2010zn}. Therefore, the 
challenge is to define a way to constrain and quantify the background,
while in parallel isolating the signal that comes from the CME. 

A first step in this direction was taken by the ALICE
Collaboration~\cite{Acharya:2017fau} using a method proposed and
developed in Ref.~\cite{Schukraft:2012ah}. This method, called Event
Shape Engineering (ESE), utilises the fluctuations of the initial
geometry and selects events with different initial system shapes,
e.g. central Pb--Pb collisions with large initial anisotropy. This
study set an upper limit of 26--33$\%$ at $95\%$  
confidence level for the CME signal contribution to the 
$\langle \cos(\varphi_{\alpha} + \varphi_{\beta} - 2\Psi_{2})
\rangle$) correlator in the 10--50$\%$ centrality interval.
The CMS~\cite{Khachatryan:2016got} and the
STAR~\cite{STAR:2019xzd} collaborations studied charge-dependent
correlations in p--Pb collisions at $\sqrt{s_{\mathrm{NN}}} = 5.02$~TeV
and in p--Au and d--Au collisions at  $\sqrt{s_{\mathrm{NN}}} = 0.2$~TeV,
respectively. In these colliding systems, one expects the CME
contribution to any charge-dependent signal to be small and the
results can thus be used to gauge the magnitude of the background in
heavy-ion collisions. Both results illustrate that these correlations
are similar to those measured in heavy-ion collisions.
First results using ESE have also been reported by the CMS
collaboration in Ref.~\cite{Sirunyan:2017quh} which set upper limits
on the CME fraction of the three-particle correlator to be 13\% and 7\%
(at 95\% confidence level) for p–Pb and Pb–Pb collisions.

In this article we report results on two-particle correlations of
different orders as well as various two-particle correlations relative
to the second, third and fourth-order symmetry planes for charged
particles in Pb--Pb collisions at $\sqrt{s_{\rm NN}} = 2.76$~and
5.02~TeV. The motivation for utilising different planes is that the
charge separation originating from the CME is expected to  
be present along the direction of the magnetic field and thus
perpendicular to the reaction plane, approximated by $\Psi_2$. Since
the third order symmetry plane $\Psi_3$ is very weakly correlated  
with $\Psi_2$~\cite{Aad:2014fla} the charge separation effect relative
to the third harmonic symmetry plane is expected to be
negligible.
First results on correlations relative to $\Psi_3$ reported
by the CMS collaboration in Ref.~\cite{Sirunyan:2017quh}  indicates
that the charge separation could be originating from the coupling of
two-particle correlations with the anisotropic flow.
In addition, contributions from
correlations induced by the CME should be strongly suppressed in the
measurements of two-particle correlations relative to $\Psi_4$, while
the background effects stemming from local charge conservation should
scale with $v_{4}$~\cite{Voloshin:2011mx}. Therefore, measurements 
of correlations relative to higher order symmetry planes are expected
to reflect mainly, if not solely,  background effects.

The article is organised as follows: Sec.~\ref{Section:ExperimentalSetup} describes briefly the 
experimental setup, while Sec.~\ref{Section:Analysis} discusses the data sample, the selection 
criteria as well as the correlators reported; these sections are followed by Sec.~\ref{Section:Systematics} and 
Sec.~\ref{Section:Results} where the estimation of the systematic uncertainties of all 
measurements and the main physics results, respectively, are presented. We conclude in 
Sec.~\ref{Section:Summary} with a summary.

\section{Experimental setup} 
\label{Section:ExperimentalSetup}

By convention in ALICE, the beam direction defines the $z$-axis, the
$x$-axis is horizontal and points towards the centre of the LHC, and
the $y$-axis is vertical and points upwards. The apparatus consists   
of a set of detectors located in the central barrel, positioned inside
a solenoidal magnet which can generate a field parallel to the beam
direction with  maximum magnitude of 0.5 T. A set of forward detectors 
completes the  experimental setup.

The main tracking devices of ALICE are the Inner Tracking System
(\ITS)~\cite{Aamodt:2008zz} and the Time Projection Chamber
(\TPC)~\cite{Alme:2010ke}. The \ITS~consists of six cylindrical
layers of silicon detectors employing three different
technologies. The two innermost layers, positioned at $r = 3.9$~cm  
and 7.6~cm, are Silicon Pixel Detectors (\SPD), followed by two layers
of Silicon Drift Detectors (\SDD) ($r = 15$~cm and 23.9~cm). Finally,
the two outermost layers are double-sided Silicon Strip Detectors  
(\SSD) at $r = 38$~cm  and 43~cm. The \TPC~surrounds the \ITS~and
provides full azimuthal coverage. The combined pseudorapidity ($\eta$)
coverage of the \ITS~and the \TPC~is $-0.9 < \eta < 0.9$.   

A set of forward detectors, the \VZERO~scintillator
arrays~\cite{Abbas:2013taa}, were used in the trigger  
logic and for the determination of the collision centrality, discussed
in the next section. The \VZERO~consists of two sub-systems, the
\VZEROA~and the \VZEROC, that are positioned on either side of the
interaction point and cover the pseudorapidity ranges of $2.8 < \eta <
5.1$ and $-3.7 < \eta < -1.7$, respectively.  Finally the Zero Degree
Calorimeters (ZDC)~\cite{Aamodt:2008zz} positioned at both positive
and negative rapidity at around 114~m away from the interaction point
were also used offline to reduce the contamination from beam-induced
background. 

A detailed description of  ALICE and its sub-detectors can be found in
Ref.~\cite{Aamodt:2008zz} and their performance in
Ref.~\cite{Abelev:2014ffa}.  
\section{Analysis details} 
\label{Section:Analysis}

\subsection{Event and track selection} 
The analysis is performed using the Pb--Pb data samples collected in 2010
and 2015 at a centre-of-mass energy per nucleon pair of $\sqrt{s_\textrm{NN}}$ =
2.76 and 5.02 TeV, respectively. The minimum bias trigger condition is
defined in the 2010 data sample by combinations of hits in the SPD and
either V0A or V0C detectors, while in 2015 the trigger required a
signal in both V0A and V0C detectors. 

An offline event selection relying on the timing information from the
V0 and the neutron ZDC is used to reject beam-gas background and
parasitic beam-beam interactions. Events are analysed if the 
$z$--coordinate of the reconstructed primary vertex $(V_{z})$ resides within
$\pm$10~cm from the nominal interaction point.  The collision 
centrality is estimated from the amplitude of the signal measured by the V0
detectors as explained in Ref.~\cite{Abelev:2013qoq}. Higher amplitude,
and hence higher particle multiplicity, corresponds to more central
(smaller impact parameter) events. The data sample is divided into
centrality classes which span 0--70$\%$ of the inelastic hadronic 
cross section, which is considered in this study. The 0--5$\%$ and
60--70\% intervals correspond to the most central and the
most peripheral collisions, respectively. 

Charged particles reconstructed using the \TPC~and the \ITS~information are
accepted for analysis within $\eta$ and $p_{\rm{T}}$ ranges of $|\eta|
< 0.8$ and $0.2 < p_{\rm{T}} < 5$~GeV/$c$, respectively. The
tracking algorithm, based on the Kalman filter~\cite{Billoir:1983mz,Billoir:1985nq}, 
starts from a collection of space points (referred to as clusters)
inside the \TPC, and provides the quality of the fit by calculating  
its $\chi^2$ value. 
The track parameters at the primary vertex are then updated using the combined
information from both the \TPC~and the \ITS~detectors. Tracks are
accepted even if the algorithm is unable to match the track
reconstructed in the \TPC~with associated \SPD~clusters (e.g.~due to
inefficiencies caused by dead channels in the \SPD~layers). In this
case, a cluster from another layer of the ITS (e.g. SDD) is used to
reconstruct the tracks.
This tracking mode will be referred to as hybrid
tracking in the rest of the text and is used as the default in this
analysis since it provides a uniform distribution in azimuthal angle ($\varphi$). 
More details about the tracking parameters and performance are
described elsewhere~\cite{Aamodt:2008zz,Abelev:2014ffa}. 
Accepted tracks are required to have at least 70 out of 159 possible
space points measured in the \TPC and a $\chi^2$ per degree
of freedom of the momentum fit per \TPC~cluster to be below 2. These  
selections reduce the contribution from short tracks, which are
unlikely to originate from the primary vertex. To further reduce the
contamination by secondary tracks from weak decays or from the
interaction with the material, only tracks within a maximum
distance of closest approach (DCA) to  
primary vertex in both the transverse plane ($\mathrm{DCA}_{\rm xy} <
2.4$~cm) and the longitudinal direction ($\mathrm{DCA}_{\rm z} < 3.2$~cm)
were considered. Moreover, if matched to ITS clusters, the tracks are
required to have at least one cluster in either of the two \SPD~layers.
These selections lead to an efficiency of about $65\%$
for primary tracks at $p_\mathrm{T} = 0.5$~GeV/$c$, which reaches  
$80\%$ above 1~GeV/$c$. The variation of these values 
between central and peripheral collisions is less than 3$\%$, and
does not change between $\sqrt{\it{s}_\mathrm{{NN}}} = $ 2.76~and 5.02~TeV.
The contamination from secondaries is about $10\%$ at
$p_{\rm{T}} = 0.2$~GeV/$c$, reaches $5\%$ at $p_{\rm{T}} = 1$~GeV/$c$
and decreases further with increasing transverse momentum.

\subsection{Analysis methodology}
\label{sec:factorization}
A way to probe the P-odd leading order coefficient $a_{1,\alpha}$ 
that reflects the magnitude of the CME is through the study of
charge-dependent two-particle correlations relative to the reaction
plane $\Psi_{\rm RP}$. The expression proposed in Ref.~\cite{Voloshin:2004vk}
is of the form  $\langle \cos(\varphi_{\alpha} + \varphi_{\beta} -
2\Psi_{\rm RP}) \rangle$ ($\alpha$ and $\beta$ being particles with the
same or opposite charges) that can probe correlations between the leading
P-odd terms for different charge combinations $\langle a_{1,\alpha} a_{1,\beta} \rangle$.  
This can be seen if one decomposes the correlator using Eq.~\ref{Eq:Fourier}

\[\langle \cos(\varphi_{\alpha} + \varphi_{\beta} - 2\Psi_{\rm RP}) \rangle = \] 

\[ \langle \cos\big[(\varphi_{\alpha} - \Psi_{\rm RP}) +
(\varphi_{\beta} - \Psi_{\rm RP})\big] \rangle = 
\langle \cos(\Delta \varphi_{\alpha} + \Delta \varphi_{\beta}) \rangle = \]

\begin{equation}
\langle \cos \Delta \varphi_{\alpha} \cos \Delta \varphi_{\beta} \rangle - 
\langle \sin \Delta \varphi_{\alpha} \sin \Delta \varphi_{\beta} \rangle = 
\langle v_{1,\alpha}v_{1,\beta} \rangle + \mathrm{B_{in}} - 
\langle a_{1,\alpha} a_{1,\beta}\rangle - \mathrm{B_{out}},  
\label{Eq:3ParticleCorrelator}
\end{equation}

where $\mathrm{B_{in}}$ and $\mathrm{B_{out}}$ represent the parity-conserving 
correlations projected onto the in- and out-of-plane directions. The terms
$\langle \cos \Delta \varphi_{\alpha} \cos \Delta \varphi_{\beta}\rangle$
and $\langle \sin \Delta \varphi_{\alpha}\sin \Delta\varphi_{\beta}\rangle$
in Eq.~\ref{Eq:3ParticleCorrelator} quantify the correlations with
respect to the in- and out-of-plane directions, respectively. The term
$\langle v_{1,\alpha}v_{1,\beta} \rangle$, i.e. the product of the
first order Fourier harmonics or directed flow, is expected to have
negligible charge dependence in the midrapidity region~\cite{Gursoy:2018yai}. In addition,
for a symmetric collision system the average directed flow at
midrapidity is zero.
A generalised form of Eq.~\ref{Eq:3ParticleCorrelator} also describing
higher harmonics is given by the mixed-harmonics correlations, which
reads  
\begin{eqnarray}
\label{eq:moments}
\gamma_{\mathrm{m,n}} = \langle \cos(\mathrm{m}\varphi_{\alpha} +
\mathrm{n}\varphi_{\beta} - \mathrm{(m + n)}\Psi_{\mathrm{|m+n|}})\rangle , 
\label{Eq:Generalised3ParticleCorrelator}
\end{eqnarray}

\noindent where $\mathrm{m}$ and $\mathrm{n}$ are integers. Setting $\mathrm{m}
= 1$ and $\mathrm{n} = 1$ (i.e. $\gamma_{1,1}$) leads to Eq.~\ref{Eq:3ParticleCorrelator}. 
The $\mathrm{|m+n|}$-th order symmetry plane angle
$\Psi_{\mathrm{|m+n|}}$ is introduced to take into account that the
overlap region of the colliding nuclei exhibits an irregular 
shape~\cite{Manly:2005zy,Bhalerao:2006tp,Alver:2008zza,Alver:2010gr,Alver:2010dn}. 
This originates from the initial density profile of nucleons participating in the collision, 
which is not isotropic and differs from one event to the other. In case of a smooth 
distribution of matter produced in the overlap zone, the angle
$\Psi_{\mathrm{|m+n|}}$ coincides with that of the reaction plane,
i.e. $\Psi_{\mathrm{|m+n|}} = \Psi_{\rm RP}$.  

In order to independently evaluate the
contributions from correlations in- and out-of-plane, one can also
measure a two-particle correlator of the form
\[ \langle \cos(\varphi_{\alpha} - \varphi_{\beta}) \rangle = 
\langle \cos\big[(\varphi_{\alpha} - \Psi_{\rm RP}) -
(\varphi_{\beta} - \Psi_{\rm RP})\big] \rangle =  
\langle \cos(\Delta \varphi_{\alpha} - \Delta \varphi_{\beta}) \rangle= \] 
\begin{eqnarray}
\langle \cos \Delta \varphi_{\alpha} \cos \Delta \varphi_{\beta} \rangle +
\langle \sin \Delta \varphi_{\alpha} \sin \Delta \varphi_{\beta} \rangle = 
\langle v_{1,\alpha}v_{1,\beta} \rangle + \mathrm{B_{in}} + 
\langle a_{1,\alpha} a_{1,\beta}\rangle + \mathrm{B_{out}},
\label{Eq:2ParticleCorrelator}
\end{eqnarray}

which corresponds to the special case of $\mathrm{m = -n}$ in
Eq.~\ref{Eq:Generalised3ParticleCorrelator}. This provides access to the 
two-particle correlations without any dependence on the symmetry
plane angle 
\begin{eqnarray}
\delta_\mathrm{m} = \langle \cos[\mathrm{m}(\varphi_{\alpha} -
\varphi_{\beta})] \rangle .
\label{Eq:Generalised2ParticleCorrelator}
\end{eqnarray}
\noindent This correlator, owing to its construction, is affected if
not dominated by non-flow contributions. Charge-dependent results for
$\delta_1$, together with the relevant measurements of $\gamma_{1,1}$ were 
first reported in Ref.~\cite{Abelev:2012pa} and made it possible to
separately quantify the magnitude of correlations in- and out-of-plane. 

In this article, we report on the charge-dependent results of four
correlators of the form of
Eq.~\ref{Eq:Generalised3ParticleCorrelator}. The first two,
$\gamma_{1,1}$ and $\gamma_{1,-3}$, probe correlations of particles
relative to the second order symmetry plane 
($\Psi_2$). The correlator $\gamma_{1,1}$ (i.e. the main correlator
used in previous studies) probes correlations of the first order P-odd
term, i.e.  $\langle a_{1,\alpha} a_{1,\beta} \rangle$ as
illustrated in Eq.~\ref{Eq:3ParticleCorrelator}, while the second is
sensitive not only to the first but also the third order coefficient,
i.e. $\langle a_{1,\alpha} a_{3,\beta} \rangle$ and
thus is sensitive to the magnitude and the shape of the CME
contribution. However, in both cases the background contributions 
from local charge conservation are expected to be significant
(see Ref.~\cite{Schlichting:2010qia,Pratt:2010zn} and the references  
therein). 

In order to evaluate the background, correlations relative to the
third and fourth order symmetry planes i.e.,  $\gamma_{1,2}$ and
$\gamma_{2,2}$, are investigated. Since the charge-separation effects
originating from the CME form relative to the second order symmetry
plane, both correlators are expected to have negligible
contribution from it. Their charge-dependent part could thus be used 
as a proxy for the background that consists of local charge conservation
scaled by the corresponding flow harmonics according to Ref.~\cite{Bzdak:2012ia} 

\begin{subequations}
\label{Eq:3ParticleCorrelatorsBackground}
 \begin{align}
\gamma_{1,1} \approx \langle \cos[(\varphi_{\alpha} - \varphi_{\beta})
   + 2(\varphi_{\beta} - \Psi_2)] \rangle \propto \delta_1 v_2, \\
\gamma_{1,2} \approx \langle \cos[(\varphi_{\alpha} - \varphi_{\beta})
   + 3(\varphi_{\beta} - \Psi_3)] \rangle \propto \delta_1 v_3 , \\
\gamma_{2,2} \approx \langle \cos[2(\varphi_{\alpha} -\varphi_{\beta})
   + 4(\varphi_{\beta} - \Psi_4)] \rangle \propto \delta_2 v_4.
 \end{align}
\end{subequations}

By taking the difference of results between opposite- and same-sign
charge combinations, denoted as $\Delta\gamma_{\mathrm{mn}}$ in the
most general form of the correlator, one can eliminate the
charge-independent part and probe the contribution from local charge
conservation modulated by the relevant flow harmonic

\begin{subequations}
\label{Eq:cmeFactorise}
 \begin{align}
   \Delta\gamma_{1,1} \approx \kappa_{2}v_{2}\Delta\delta_{1},\\
   \Delta\gamma_{1,2} \approx \kappa_{3}v_{3}\Delta\delta_{1}, \\
   \Delta\gamma_{2,2} \approx  \kappa_{4}v_{4}\Delta\delta_{2},
 \end{align}
\end{subequations}

\noindent where $\kappa_{n}$ is a proportionality constant.
Using Eqs.~\ref{Eq:cmeFactorise}, one can thus estimate the
contribution of the background in the charge-dependent CME sensitive
correlator $\Delta\gamma_{1,1}$ using the results of e.g. $\Delta
\gamma_{1,2}$ according to  

\begin{equation}
\label{Eq:dataBkgC112}
\Delta\gamma_{1,1}^{\textrm{Bkg}} \approx  \Delta\gamma_{1,2} \times
\frac{v_{2}}{v_{3}} \frac{\kappa_{2}}{\kappa_{3}}.
\end{equation}

Equation~\ref{Eq:dataBkgC112} serves as a tool to disentangle the CME   
contribution from the background, provided the parameter
$\kappa_{2}/\kappa_{3}$ is estimated. In
Ref.~\cite{Khachatryan:2016got}  it was argued that the magnitude of   
these $\kappa_{n}$ terms depends on the kinematic ranges
(e.g. detector acceptance, event and particle selection criteria).
Although $\kappa_{n}$ may have dependency on $p_{\rm T}$ and $\eta$,  
we have have ignored such dependency and assumed a constant magnitude 
of the ratio $\kappa_{2}/\kappa_{3}$ for the full kinematic range. 
In Ref.~\cite{Khachatryan:2016got}, it was suggested that one can
assume that $\kappa_{2} \approx \kappa_{3}$ if the same kinematic 
conditions are used to calculate $\Delta\gamma_{\mathrm{m,n}}$ 
within the same experimental setup. In this article, we also
investigate the relationship between $\kappa_{2}$ and $\kappa_{3}$ 
using two approaches: a blast wave~\cite{Retiere:2003kf} inspired
model that incorporates effects of local charge conservation and the
results of A Multi Phase Transport model
(AMPT)~\cite{Zhang:1999bd,Lin:2000cx,Lin:2004en}, both discussed  
in detail in the Results section.

\subsubsection{The event-plane method}
To evaluate the correlations experimentally, the event-plane
method~\cite{Poskanzer:1998yz,Voloshin:2008dg} is used. In this
method, the event plane angle is reconstructed from the
azimuthal distribution of the particles produced in a collision. The
event plane angle of k-th order (where $\mathrm{k} = |\mathrm{m -
  n}|$) $\Psi_{\mathrm{k,EP}}$ is estimated according to  

\begin{equation}
\Psi_{\mathrm{k,EP}} = \tan^{-1}\frac{Q_{\mathrm{k,y}}}{Q_{\mathrm{k,x}}},
\end{equation}

\noindent where $Q_{\mathrm{k,x}}$ and $Q_{\mathrm{k,y}}$ are the x-
and y-components of the Q-vector, calculated as

\begin{equation}
Q_{\mathrm{k,x}} = \sum_{i=1}^{M}
w_{i}(p_{\mathrm{T}},\eta,\varphi,V_{z})\cos(\mathrm{k}\varphi_{i}),
 \ {\rm and\ } 
 Q_{\mathrm{k,y}} = \sum_{i=1}^{M}
w_{i}(p_{\mathrm{T}},\eta,\varphi,V_{z})\sin(\mathrm{k}\varphi_{i}). 
\label{Eq:QVectorComponents}
\end{equation}
In Eq.~\ref{Eq:QVectorComponents}, $\varphi_{i}$ corresponds to the
azimuthal angle of the $i$-th track in an event with multiplicity
$M$. The factors $w_{i}(p_{\mathrm{T}},\eta,\varphi,V_{z})$ are weights
applied on every track in the construction of the Q-vectors, in order
to correct for non-uniform reconstruction efficiency and
acceptance. They are calculated as a function of the transverse
momentum, pseudorapidity and azimuthal
angle of particles for different $V_{z}$ values of the
primary vertex. 

To reduce the contributions from short range effects not related to
the common symmetry planes (i.e. non-flow), a subevent plane
technique~\cite{Poskanzer:1998yz,Voloshin:2008dg} is implemented. Each 
event is divided  into two subevents ``$A$'' and ``$B$'', covering the
ranges $-0.8<\eta<0$ and $0<\eta<0.8$, respectively, and the two
subevent plane angles, namely $\Psi_{\mathrm{k},A}$ and $\Psi_{\mathrm{k},B}$
are calculated using charged particles. The correlators of
Eq.~\ref{Eq:Generalised3ParticleCorrelator} are then calculated as 
\begin{equation}
\gamma_{\mathrm{mn}} = \frac{\langle
  \cos[\mathrm{m}\varphi_{\alpha} + \mathrm{n}\varphi_{\beta} -
  \mathrm{(m + n)}\Psi_{\mathrm{|m+n|},\mathrm{EP}}]
  \rangle}{R(\Psi_{\mathrm{|m+n|},\mathrm{EP}})},  
\end{equation}
\noindent where $\alpha$ and $\beta$ correspond to any two charged
particles within $-0.8<\eta<0.8$, and $\Psi_{\mathrm{|m+n|},\mathrm{EP}}$
corresponds to subevent plane $\Psi_{\mathrm{k},A}$ (or
$\Psi_{\mathrm{k},B}$ for systematic studies). 
Particles $\alpha$ or $\beta$ (or both) were excluded from the
determination of event plane if they were from the same
$\eta$ window as the one used to calculate $\Psi_{\mathrm{k},A}$ or
$\Psi_{\mathrm{k},B}$.  

The event plane resolution $R(\Psi_{\mathrm{|m+n|},\mathrm{EP}})$ is
given by     
\begin{equation}
 R(\Psi_{\mathrm{|m+n|},\mathrm{EP}}) = \sqrt{ \cos[\mathrm{|m+n|}
   (\Psi_{\mathrm{|m+n|},A} - \Psi_{\mathrm{|m+n|},B} ) ] }. 
\end{equation}
The amount of non-flow correlations in the results of both same- (SS)
and opposite-sign (OS) charge combinations could also depend on the
longitudinal position of the detector used for the estimation of
$\Psi_{k}$. However, it was checked that the charge-dependent
differences, i.e. OS-SS are not affected by this choice as these
non-flow contributions (approximately) cancel out in the subtraction. 

\section{Systematic uncertainties} 
\label{Section:Systematics}
The systematic uncertainties in all measurements presented in this
article were estimated by varying the event and track selection
criteria as well as by studying the detector effects with Monte Carlo
(MC) simulations. The contributions from different sources, described
below, were extracted from the difference for the results of each
correlator obtained with the primary selection criteria and the ones
after the relevant variation was applied. All sources with a
difference between the results larger than 1$\sigma$ were then  
added in quadrature to form the final value of the systematic
uncertainty (for each data point), where  $\sigma$
is the uncertainty of the difference between the default results and
the ones obtained from the variation of the selection criteria, taking
into account the degree of their correlation~\cite{Barlow:2002yb}. 

Table~\ref{Tab:SystematicSources} summarises the sources and the
variations that were tested. In particular, the systematic uncertainty
originating from the selection of the z position of the primary vertex 
was investigated by changing this selection  from $\pm$10~cm down to
$\pm$8~cm. In order to estimate the contribution to the results from
the choice of the detector used as centrality estimator, the analysis
was performed using the number of hits in the second layer of the \SPD
 instead of the amplitude of the \VZERO~detector.
Furthermore, data samples recorded with different magnetic field 
configurations for the solenoid magnet were analysed separately. 
The contribution of residual pile-up events to the results was
estimated by analysing independently the high and low interaction
rate samples. Finally, the results were obtained
separately by calculating the event plane from different
pseudorapidity ranges within the \TPC~acceptance.
The systematic uncertainty in the extraction of the CME fraction when
using different event plane angles within the \TPC~acceptance for the
highest LHC energy was estimated considering runs with low beam
intensity where the distortions in the \TPC~are negligible. 

In parallel, to investigate any potential bias originating from the quality 
of the tracks used in the analysis, the number of space points
measured in the \TPC~was varied from 70 (default) up to 100 out of 159  
maximum points that a track can have. The contribution stemming from
secondary tracks, either from weak decays or from the interaction of
particles with the detector material, was investigated by tightening  
the selection on the DCA in the longitudinal direction as well as
in the transverse plane. Finally, 
another tracking mode that relies on the combination of the \TPC~and
the \ITS~detectors, henceforth called global tracking, with tighter selection
criteria in addition to requirements for clusters in the \SPD~or the
\SDD~detectors was used. In this case, a stricter transverse momentum 
dependent requirement in the value of the DCA in the transverse
plane resulted in reducing even further the amount of secondary particles 
in the track sample. The resulting contamination from secondaries is less than 
2--3\%~for the entire $p_{\rm{T}}$ range.

For each variation, new correction maps for detector inefficiencies
and non-uniform acceptance were extracted using MC data
samples and collision data.

\begin{table}[h!]
\caption{List of the selection criteria and the corresponding
  variations used for the estimation of the systematic uncertainties.}
\begin{center}
\begin{tabular}{ |l|c|c| }
\hline
 (No.) Source & Default Value & Variations \\ 
\hline
 (1)  Primary $V_{z}$ & $\pm$10 cm & $\pm$8 cm \\  
\hline
 (2)  Centrality Estimator & V0 amplitude & SPD cluster  \\
\hline
 (3) Magnetic field polarity & Combined & Positive, Negative \\
\hline
 (4) Event plane & $\Psi_{\rm k,-0.8<\eta<0}$ & $\Psi_{\rm k,0<\eta<0.8}$ \\
\hline 
 (5)  Residual Pile-Up & High Intensity data  & Low Intensity data \\
\hline 
 (6) TPC space points & 70 & 100 \\
\hline 
 (7)  DCA$_{xy}$ (DCA$_{z}$) &  2.4 (3.2) cm  & 2.0 (2.0) cm \\ 
\hline 
 (8) Tracking Algorithm & Hybrid  & Global \\
\hline 
 (9) Charge Combination & $``++"$ and $``--"$ combined & $``++"$ or $``--"$ \\
\hline
\end{tabular}
\label{Tab:SystematicSources}
\end{center}
\end{table}

Tables~\ref{Tab:SystematicValues3p}~and~\ref{Tab:SystematicValues2p}~summarise
the maximum magnitude, over all centrality intervals, of the
systematic uncertainties from each individual source for all
correlators presented in this article. The uncertainties are reported
separately for the results for same-sign (SS), opposite-sign (OS)
and the difference between opposite- and same-sign (OS-SS) pairs.
The uncertainties for the results of the various $\gamma_{m,n}$ are
reported without the common factor of $\times10^{-5}$. 

Throughout the centrality intervals
reported in this article, the magnitude of $\gamma_{1,1}$  
correlator varies between -2.4 to -40 for SS pair, -1.2 to  29 for OS
pair and 1.2 to 68 for OS-SS. The values of $\gamma_{1,-3}$ vary
between -2.1 to 38 for SS pair, -0.67 to 68 for OS pair and  1.4 to 30
for OS-SS. The magnitude of $\gamma_{1,2}$ covers the range between
-2.5 to 140 for SS pair, -1.7 to 180 for OS pair and  0.71 to 3.7 for
OS-SS. Finally, the results for $\gamma_{2,2}$ vary between 0.01 to
14.7 for SS and OS pair while being between 0.25 and 19 for OS-SS.

The two-particle correlators of the form $\delta_m$ are an order of
magnitude larger than the three-particle correlators. Therefore, the
values mentioned in the following have an exponent of
$\times10^{-4}$. The magnitude of $\delta_{1}$ varies between  2.9 to
23.5 for SS pair, 5.6 to 49 for OS pair and 2.7 to 26.2 for OS-SS. The
values of $\delta_{2}$ spans the range between 8.2 to 97 for SS pair,
9.5 to 102 for OS pair and 1.31 to 5.2 for OS-SS. The magnitude of 
$\delta_{3}$ varies between 4.5 to 16 for SS pair, 4.8 to 15 for OS
pair and -1.3 to 0.79 for OS-SS. Finally, the results for $\delta_{4}$
varies between 1.58  to 9.4 for SS pair, 1.6  to 6.8 for OS pair and
-2.5 to 0.78 for OS-SS.

\begin{table}[h]
\small
\caption{Maximum systematic uncertainty (absolute value) over all
  centrality intervals on $\gamma_{\mathrm{mn}}$ from individual
  sources (see Tab.~\ref{Tab:SystematicSources} for an explanation of
  each source). The ranges are similar for both energies.} 
\vspace{-0.5cm}
\begin{center}
\begin{tabular}{|c|c|c|c|c|c|c|c|c|c|c|c|c|}
\hline
\multirow{3}{*}{Sources} & \multicolumn{3}{c|}{$\gamma_{1,1}$ ($\times 10^{-5}$)} & %
    \multicolumn{3}{c|}{$\gamma_{1,2}$ ($\times 10^{-5}$)} &
    \multicolumn{3}{c|}{$\gamma_{1,-3}$  ($\times 10^{-5}$)}
     & \multicolumn{3}{c|}{$\gamma_{2,2}$  ($\times 10^{-5}$)} \\
\cline{2-13}
\cline{2-13}
 & SS & OS & OS-SS & SS & OS & OS-SS & SS & OS & OS-SS & SS & OS & OS-SS \\
\hline
 (1)  & 0.26 & 1.4 & 0.027 & 1.1 & 0.12 & 1.9 & 0.13 & 0.15 & 0.095 & 0.035 & 0.1 & 0.02 \\ \hline
 (2)  & 2.5   & 6.1 & 6 & 4.5 & 9.9 & 1.8 & 4.2 & 3.2 & 1.2 & 8.6 & 8.6 & 0.27 \\ \hline
 (3)  & 0.86 & 0.65 & 0.1 & 0.83 & 0.84 & 0.024 & 0.34 & 0.54 &  0.04 & 1.4 & 0.98 & 0.36 \\ \hline
 (4)  & 1.62 & 1.81 & 1.6 & 1.7 & 1.83 & 4.51 & 0.86 & 0.64 & 0.15 & 1.78 & 2.1 & 6.8 \\ \hline
 (5)  & 4.0 & 3.9 & 0.58 & 7.2 & 3.9 & 3.6 & 0.73 & 0.83 & 0.2  & 4.4 & 4.0 & 11 \\ \hline
 (6)  & 0.1 & 0.89 & 0.065 & 1.1 & 1.4 & 1.3 & 0.21 & 0.22 & 0.032 & 4.1 & 3.9 & 0.12 \\ \hline
 (7)  & 0.011 & 0.032 & 0.001 & 0.05 & 0.06 & 2.1 & 0.008 & 0.024 & 0.025 & 0.017 & 0.28 & 0.03\\ \hline
 (8)  & 0.045& 0.049 & 0.16 & 0.67 & 1.3 & 0.08 & 0.07 & 0.06 & 0.17 & 5.7  & 0.17 & 0.1 \\ \hline
 (9)  & 0.55 & - & 0.55 & 0.26 & - & 0.26 & 0.23 & - & 0.23 & 21 & - & 21 \\ \hline
\end{tabular}
\label{Tab:SystematicValues3p}
\end{center}
\end{table}

\begin{table}[h]
\small
\caption{Maximum systematic uncertainty (absolute value) over all
  centrality intervals on $\delta_{\mathrm{m}}$ from individual
  sources (see Tab.~\ref{Tab:SystematicSources} for an explanation of
  each source). The ranges are similar for both energies.} 
\begin{center}
\begin{tabular}{|c|c|c|c|c|c|c|c|c|c|c|c|c|}
\hline
\multirow{3}{*}{Sources} & \multicolumn{3}{c|}{$\delta_{1}$ ($\times 10^{-4}$)} & %
    \multicolumn{3}{c|}{$\delta_{2}$ ($\times 10^{-4}$)} &
    \multicolumn{3}{c|}{$\delta_{3}$  ($\times 10^{-4}$)}
     & \multicolumn{3}{c|}{$\delta_{4}$  ($\times 10^{-4}$)} \\
\cline{2-13}
\cline{2-13}
 & SS & OS & OS-SS & SS & OS & OS-SS & SS & OS & OS-SS & SS & OS & OS-SS \\ \hline
 (1) & 1.7 & 1.8 & 0.01& 0.12 & 0.13 & 0.03 & 0.08 & 0.11 & 0.03 & 0.06 & 0.05 &  0.01 \\ \hline
 (2) &  2.0  & 3.6 & 1.6 & 0.65 & 0.66 & 0.019 & 0.5 & 0.11 & 0.34 & 0.43 & 0.17 & 0.27 \\ \hline
 (3) & 0.86 & 1.0 & 0.35 & 1.6 & 1.4 & 0.25 & 0.64 & 0.22 & 0.33 & 0.52 & 0.02 & 0.51 \\ \hline
 (5)  & 1.2 & 0.91 & 1.5 & 1.4 & 0.99 & 0.029 & 0.89 & 0.38 & 0.33 & 0.61 & 0.13 & 0.022 \\ \hline
 (6)  & 0.014 & 2.2 & 1.4 & 5.5 & 5.5& 1.1 & 0.26 & 1.1 & 1.1 & 0.94 & 2.2 & 0.12 \\ \hline
 (7) & 0.056 & 0.15 & 0.01  & 0.08 & 0.14 & 0.01 & 0.05 & 0.02 & 0.02 & 0.14 & 0.02 & 0.07 \\ \hline
 (8) & 2.5 & 1.8 & 0.74 & 0.92 & 0.29 &  1.2 & 1.2 & 0.53 & 1.8 & 0.68 & 0.67  & 1.4 \\ \hline
 (9) & 2.6  & - & 2.6 & 1.6 & - & 1.6 & 1.7 & - & 1.7 & 0.37 & - & 0.37 \\ \hline

\end{tabular}
\label{Tab:SystematicValues2p}
\end{center}
\end{table}
\section{Results} 
\label{Section:Results}

\begin{figure}[t!b]
    \begin{center}
    \includegraphics[width = 0.85\textwidth]{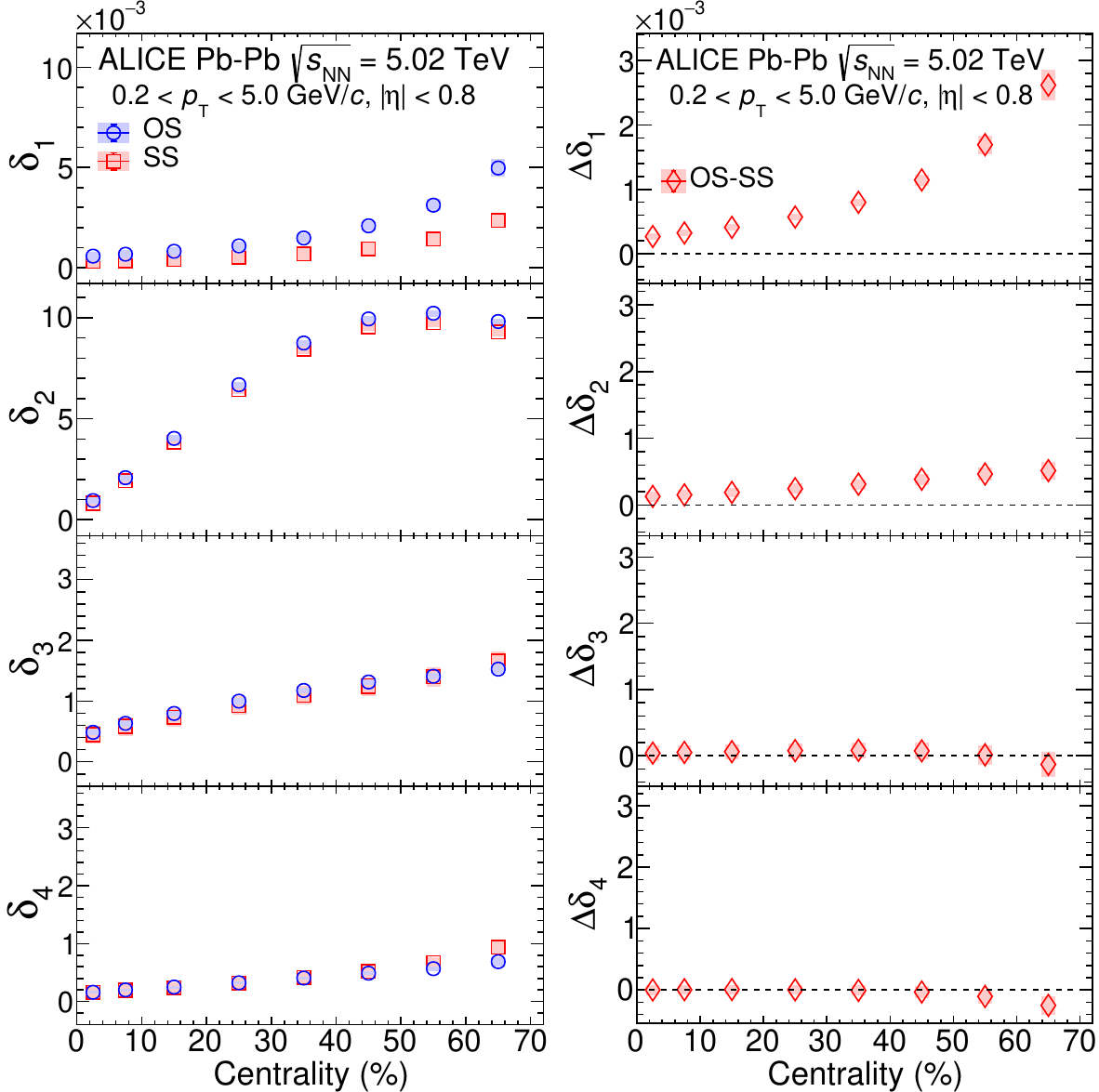}
    \end{center}
    \caption{(Left panel): The centrality dependence of $\delta_1$, $\delta_2$, $\delta_3$, and 
    $\delta_4$ for pairs of particles of opposite (OS) and same (SS) sign measured in Pb--Pb 
    collisions at $\sqrt{s_\mathrm{{\rm NN}}} = 5.02$~TeV. (Right panel): The charge-dependent 
    differences, $\Delta \delta_{\mathrm{n}}$ for $n$~=~1,~2,~3~and~4, as a function of collision 
    centrality. The statistical uncertainties for some data points
    are smaller than the marker size. The systematic uncertainties of
    each data point are represented by the shaded boxes.}  
    \label{fig:Int2pcorr}
\end{figure}

The measurements of two-particle correlators
(Eq.~\ref{Eq:Generalised2ParticleCorrelator}) are presented in
Fig.~\ref{fig:Int2pcorr}. Each data point on this figure and in the
rest of the article is drawn with the relevant statistical (vertical
lines) and systematic uncertainties (shaded boxes).The plots in the
left panel of Fig.~\ref{fig:Int2pcorr} present the centrality
dependence of $\delta_{\mathrm{m}}$ for
$\mathrm{m}$~=~1,~2,~3~and~4 for opposite (OS) and same (SS) sign
pairs. The charge-dependent differences of
every correlator, denoted by $\Delta \delta_1$, $\Delta \delta_2$,
$\Delta \delta_3$, and $\Delta \delta_4$ as a function of collision
centrality are  presented in the right panel of
Fig.~\ref{fig:Int2pcorr}. These charge-dependent two-particle
correlators (Eq.~\ref{Eq:Generalised2ParticleCorrelator}) are
primarily dominated by background effects  (see discussion in
Section~\ref{Section:Analysis}) and can thus be used to constrain the
background in the CME sensitive correlator $\gamma_{1,1}$. 
The first harmonic correlator, $\delta_1$, exhibits a significant
charge-dependent difference. This correlator is related to the
balance function also studied at the LHC~\cite{Abelev:2013csa,Adam:2015gda}. 
The present results are qualitatively consistent with the ones
in Refs.~\cite{Abelev:2013csa} and~\cite{Adam:2015gda},
i.e. oppositely charged particles are more tightly correlated in
central events resulting in a narrowing of the
balance function width in $\Delta \varphi$ and thus in a smaller value
of $\delta_1$ for central events compared to peripheral Pb--Pb collisions.
For higher harmonics, the charge-dependent
differences become progressively smaller and are compatible with zero
(up to centrality $\le$ 60\%) with a hint of negative $\Delta\delta_4$
for the most peripheral events.   

\begin{figure}[t!b]
    \begin{center}
    \includegraphics[width = 0.75\textwidth]{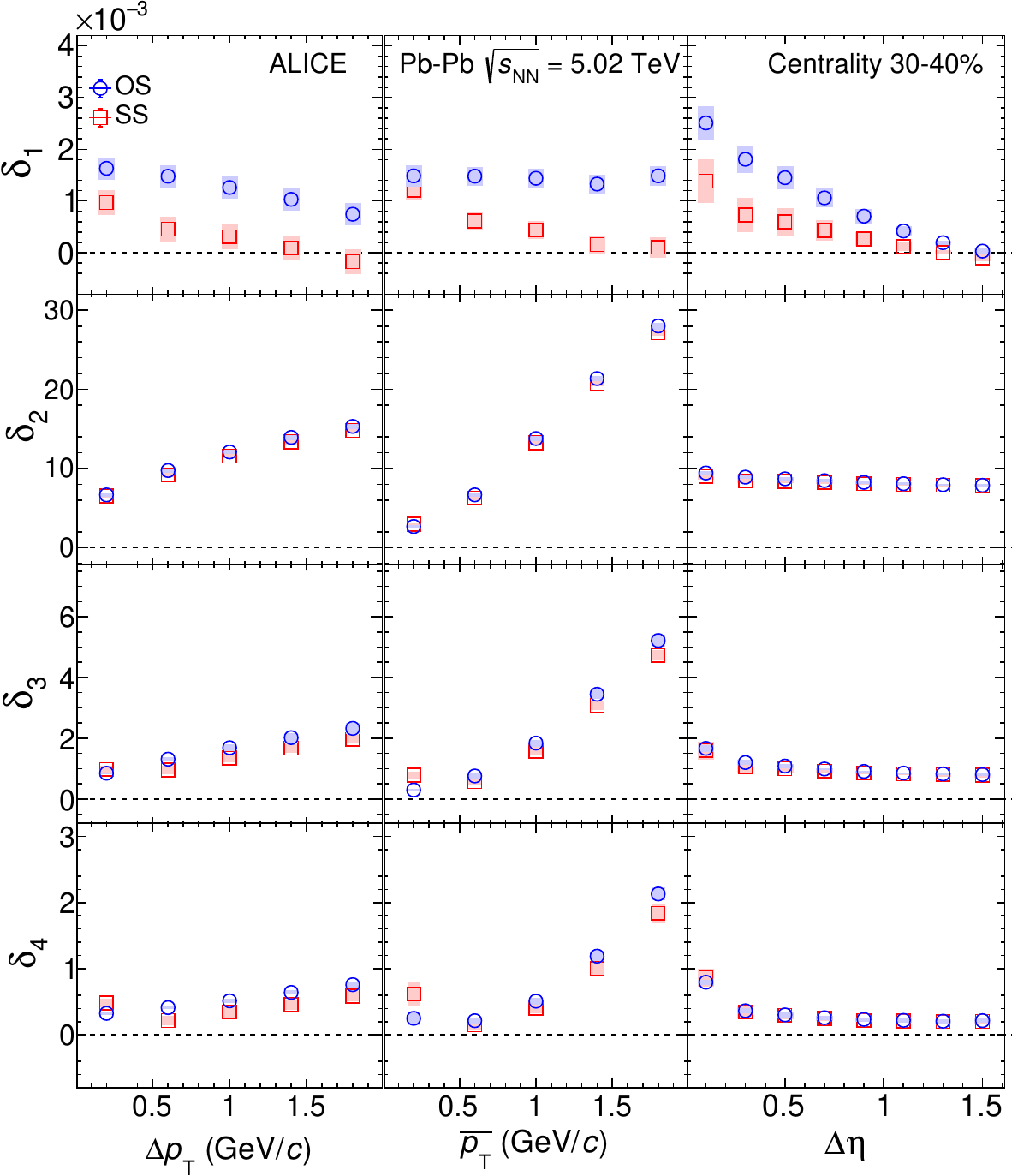}
    \end{center}
    \caption{The dependence of $\delta_1$, $\delta_2$, $\delta_3$ and
      $\delta_4$ on the transverse momentum difference
      $\Delta p_{\mathrm{T}} = |p_{\mathrm{T},\alpha} - p_{\mathrm{T},\beta}|$  
    (left panel), the average transverse momentum 
    $\overline{p}_{\mathrm{T}} = (p_{\mathrm{T},\alpha} + p_{\mathrm{T},\beta})/2$ 
    (middle panel) and the pseudorapidity difference 
    $\Delta\eta = |\eta_{\alpha} - \eta_{\beta}|$ (right panel) of
    the pair. The results for both opposite (circles) and same sign (squares) particle 
    pairs are reported for one indicative centrality interval
    (30--40$\%$) of Pb--Pb collisions at $\sqrt{s_\mathrm{{\rm NN}}}
    = 5.02$~TeV. } 
    \label{fig:Diff2pcorr}
\end{figure}

The two-particle correlators were also studied in a more differential
way, namely as a function of the transverse momentum
difference $\Delta p_{\mathrm{T}} = |p_{\mathrm{T},\alpha} - p_{\mathrm{T},\beta}|$, 
the average transverse momentum 
$\overline{p}_{\mathrm{T}} = (p_{\mathrm{T},\alpha} + p_{\mathrm{T},\beta})/2$ 
and the pseudorapidity difference 
$\Delta\eta = |\eta_{\alpha} - \eta_{\beta}|$ of the pair.

The dependence of $\delta_1$, $\delta_2$, $\delta_3$ and $\delta_4$ on
these variables for one indicative centrality interval (30--40$\%$) 
is shown in Fig.~\ref{fig:Diff2pcorr} for Pb--Pb collisions at 
$\sqrt{s_\mathrm{{\rm NN}}} = 5.02$~TeV. 
For the first harmonic correlator, $\delta_1$, the correlations
between particles of opposite charges have larger magnitude compared
with the ones  for same charge particles. The absolute differences do
not show any significant $\Delta p_{\mathrm{T}}$ dependence, however
they do increase with increasing $\overline{p}_{\mathrm{T}}$ of the
pair. Finally, there is a significant charge-dependent difference of
$\delta_1$, which decreases with increasing $\Delta \eta$, consistent
with what is also reported in Ref.~\cite{Abelev:2013csa,Adam:2015gda}. For
higher  harmonics, no significant difference is observed. For other
centralities the results look qualitatively similar.

\begin{figure}[t!b]
    \begin{center}
    \includegraphics[width = 0.85\textwidth]{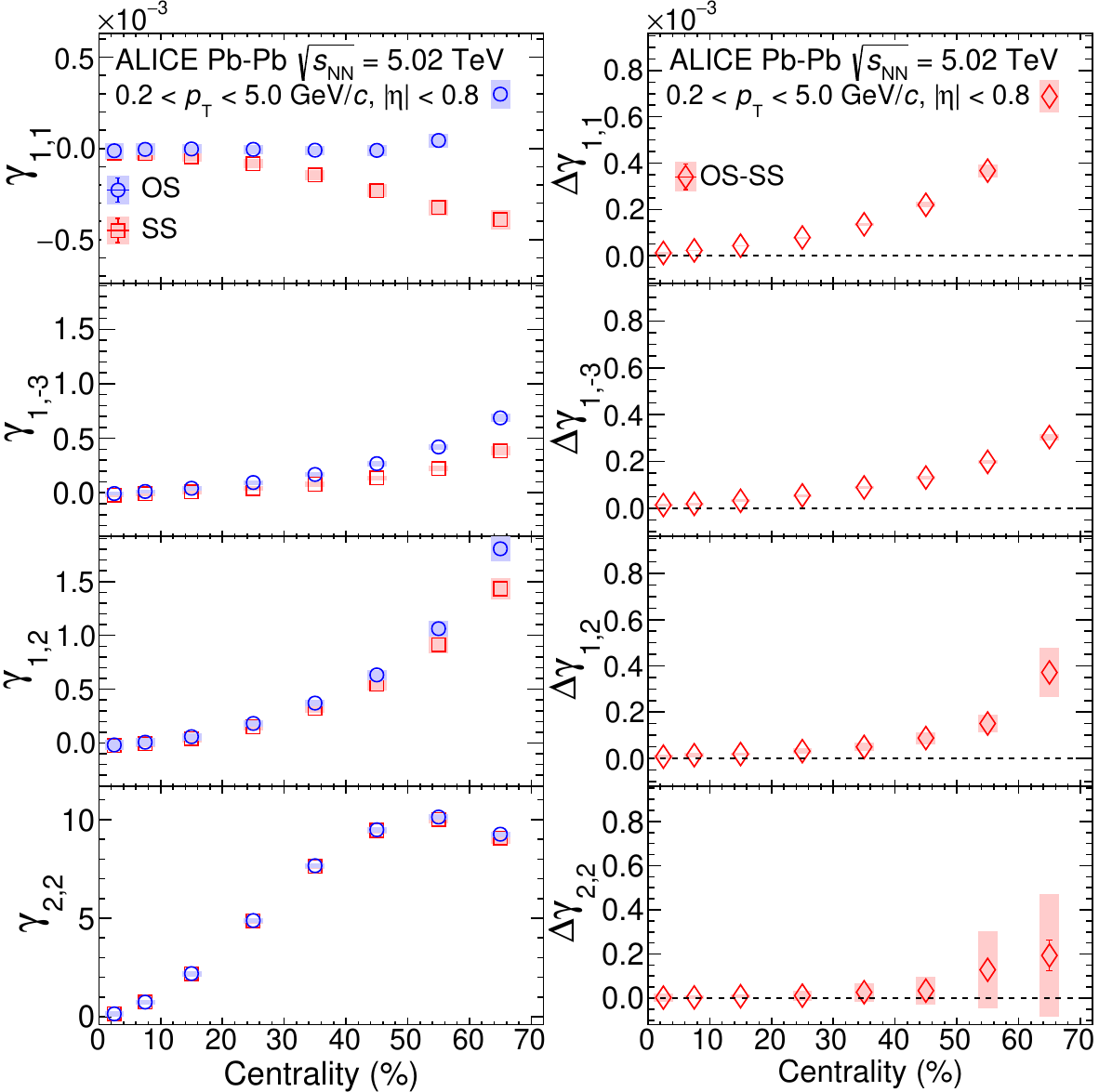}
    \end{center}
    \caption{(Left panel): The centrality dependence of $\gamma_{1,1}$, $\gamma_{1,-3}$, 
    $\gamma_{1,2}$ and $\gamma_{2,2}$ for pairs of particles of opposite (OS) and same 
    (SS) sign measured in Pb--Pb collisions at $\sqrt{s_\mathrm{{\rm NN}}} = 5.02$~TeV. (Right 
   panel): The charge-dependent differences $\Delta \gamma_{1,1}$, $\Delta \gamma_{1,-3}$, 
    $\Delta \gamma_{1,2}$ and $\Delta\gamma_{2,2}$ as a function of collision centrality.}    
    \label{fig:Int3pcorr}
\end{figure}

The measurements of integrated two-particle correlators relative to
various order symmetry planes
(Eq.~\ref{Eq:Generalised3ParticleCorrelator}) in Pb--Pb collisions at  
$\sqrt{s_\mathrm{{\rm NN}}} = 5.02$~TeV are presented in
Fig.~\ref{fig:Int3pcorr}. The left panel presents the centrality
dependence of $\gamma_{1,1}$, $\gamma_{1,-3}$, $\gamma_{1,2}$ and
$\gamma_{2,2}$. Results for different charge combinations, i.e. OS and
SS pairs are also presented here. The right panel of the same figure
presents the centrality dependence of the charge-dependent
differences, i.e. OS-SS. A significant charge-dependent  
magnitude for $\gamma_{1,1}$ is observed that increases when moving to
more peripheral collisions. In particular, the magnitude of the
same-sign correlations becomes progressively more negative, while
correlations of oppositely charged particles are very close to zero
and their magnitude turns positive for peripheral Pb--Pb events.
A significant charge-dependent difference that increases for peripheral
centrality intervals is also observed for $\gamma_{1,-3}$. Both
correlators, as discussed in Sec.~\ref{Section:Analysis}, probe
correlations between either the first order P-odd term of the form
$\langle a_{1,\alpha} a_{1,\beta} \rangle$ or between the  
first and the third order coefficient $\langle a_{1,\alpha}
a_{3,\beta} \rangle$. They are thus sensitive to contributions from the CME.

The centrality dependence of $\gamma_{1,2}$ for SS and OS pairs 
and their difference also demonstrate a significant
charge dependence which increases for more peripheral
events.
Correlations of particles relative to the third order symmetry
plane are expected to probe solely the background scaled by the third
order flow harmonic ($v_{3}$) as expressed in Eqs.~\ref{Eq:3ParticleCorrelatorsBackground}. 
Hence these results indicate that the effects of local charge
conservation coupled with ${v}_3$ can induce differences in
correlations between different charges. Finally, correlations of
particles with different charge relative to the fourth order symmetry
plane, as quantified by $\gamma_{2,2}$, do not exhibit any
significant charge dependence within the current level of statistical
and systematic uncertainties.   

\begin{figure}[t!b]
    \begin{center}
    \includegraphics[width = 0.75\textwidth]{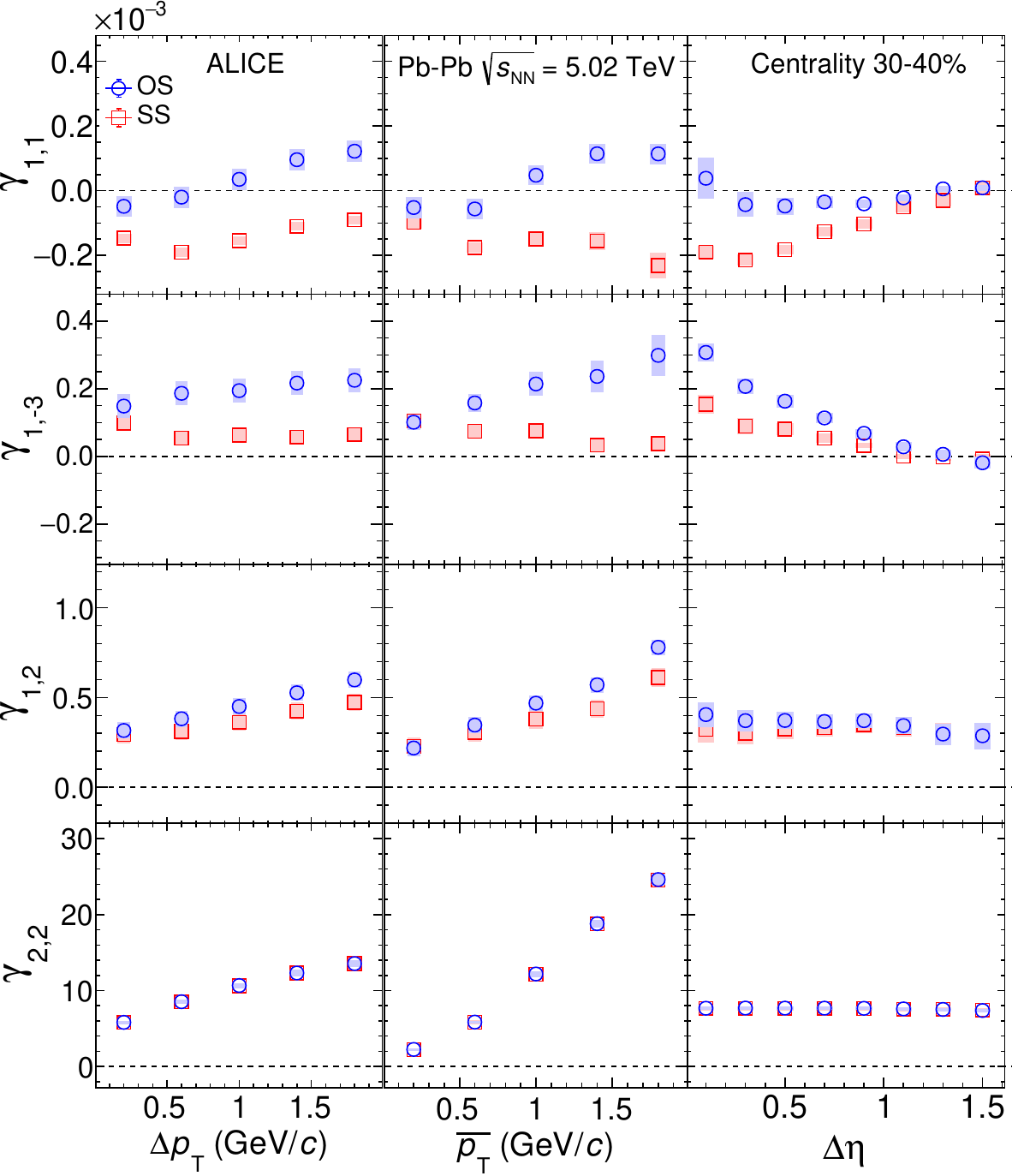}
    \end{center}
    \caption{The dependence of $\gamma_{1,1}$, $\gamma_{1,-3}$,
      $\gamma_{1,2}$ and $\gamma_{2,2}$ on the transverse
      momentum difference $\Delta p_{\mathrm{T}} =
      |p_{\mathrm{T},\alpha} - p_{\mathrm{T},\beta}|$ (left panel),
      the average transverse momentum $\overline{p}_{\mathrm{T}} =
      (p_{\mathrm{T},\alpha} + p_{\mathrm{T},\beta})/2$ (middle
      panel) and the pseudorapidity difference $\Delta\eta =
      |\eta_{\alpha} - \eta_{\beta}|$ (right panel) of the pair. The
      results for both opposite and same sign particle pairs are
      reported for one indicative centrality interval
      (30--40$\%$) of Pb--Pb collisions at
      $\sqrt{s_\mathrm{{\rm NN}}} = 5.02$~TeV.} 
    \label{fig:Diff3pcorr}
\end{figure}

As in the case of the two-particle correlators, $\delta_{\mathrm{m}}$, 
also the $\gamma_{\mathrm{m,n}}$ were studied in a differential way,
namely as a function of  $\Delta p_{\mathrm{T}}$, $\overline{p}_{\mathrm{T}}$  
and $\Delta\eta$. The results are presented in Fig.~\ref{fig:Diff3pcorr} 
for the same representative centrality interval as before (30--40$\%$) for both
OS and SS. It is seen that, with the exception of $\gamma_{2,2}$, the  
magnitude of correlations for OS pairs is greater than the one of SS
for nearly the full range of $\Delta p_{\mathrm{T}}$, $\overline{p}_{\mathrm{T}}$ 
and $\Delta\eta$ presented in this article. The results for OS and SS
are compatible within the current level of statistical and systematic
uncertainties for $\gamma_{2,2}$. 

The correlations of particles with different charge for both
$\gamma_{1,1}$ and $\gamma_{1,-3}$, i.e. the two correlators that are
sensitive to different orders of the CME, have a range that extends up
to one unit of $\Delta\eta$. 
Both OS and SS correlations have a similar trend as a function of
$\Delta p_{\mathrm{T}}$ and $\Delta \eta$, however they exhibit
different behaviour as a function of $\overline{p}_{\mathrm{T}}$. 
On the other hand, the correlators that are solely sensitive to the
background, i.e. $\gamma_{1,2}$ and $\gamma_{2,2}$, exhibit an
increasing trend as a function of both $\Delta p_{\mathrm{T}}$ and  
$\overline{p}_{\mathrm{T}}$. This trend has a mild charge dependence for
$\gamma_{1,2}$ that increases with increasing $\Delta
p_{\mathrm{T}}$ and $\overline{p}_{\mathrm{T}}$, but not for
$\gamma_{2,2}$. Both $\gamma_{1,2}$ and $\gamma_{2,2}$ have a range that extends up to $\Delta \eta =
1.6$ without any significant dependence on $\Delta\eta$.

\begin{figure}[t!b]
    \begin{center}
    \includegraphics[width =
       1.0\textwidth]{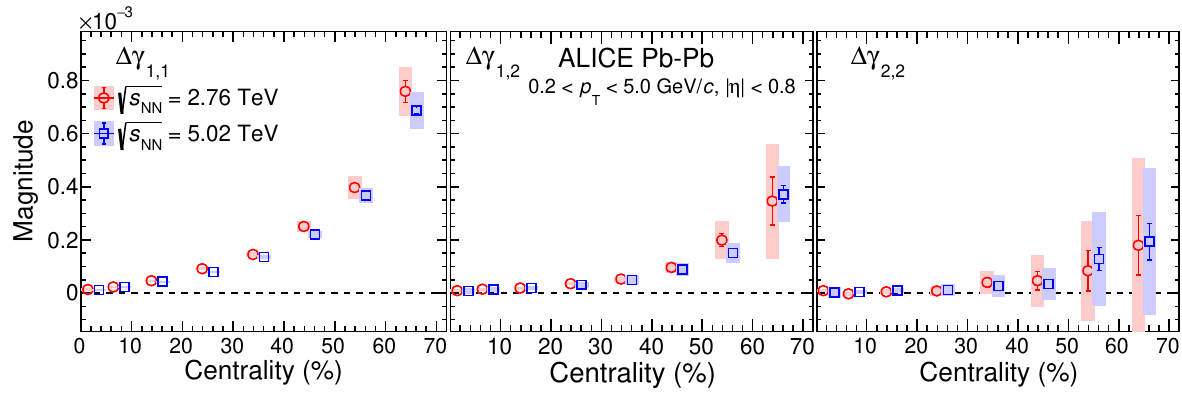}
    \end{center}
    \caption{The dependence of $\Delta \gamma_{1,1}$, $\Delta
      \gamma_{1,2}$ and $\Delta \gamma_{2,2}$ on centrality,
      measured in Pb--Pb collisions at $\sqrt{s_\mathrm{{\rm NN}}} =
      2.76$~and 5.02~TeV. The data points for $\sqrt{s_\mathrm{{\rm NN}}} =
      5.02$~TeV are shifted along the horizontal axis for better visibility.} 
    \label{fig:DeltaGammaRun1Run2}
\end{figure}

Finally, the charge-dependent differences of the correlators
$\gamma_{1,1}$, $\gamma_{1,2}$ and $\gamma_{2,2}$ were also studied in
Pb--Pb collisions at $\sqrt{s_\mathrm{{\rm NN}}} = 2.76$~TeV. 
The centrality dependence of $\Delta \gamma_{1,1}$, $\Delta
\gamma_{1,2}$ and $\Delta \gamma_{2,2}$ is presented in
Fig.~\ref{fig:DeltaGammaRun1Run2} in comparison with the results
obtained in Pb--Pb collisions at $\sqrt{s_\mathrm{{\rm NN}}} = 5.02$~TeV. 
None of the correlators exhibit any significant differences between
the two energies, within the current level of uncertainties.  
This could be explained considering that there is no significant
energy dependence in the effects that constitute the background to
these measurements (i.e. local charge conservation coupled to
different flow harmonic modulations). 
Preliminary studies indicate that the correlations between balancing
charges, as reflected in the width of the balance function, do not
exhibit any significant dependence on collision energy.
The values of $v_2$, $v_3$ and $v_4$ in $\sqrt{s_\mathrm{{\rm NN}}} =
5.02$~TeV are between 2 to 20\% higher than the values at
$\sqrt{s_\mathrm{{\rm NN}}} = 2.76$~TeV~\cite{Acharya:2018lmh}.
However, the corresponding change in the background contribution to
the $\gamma_{\rm m,n}$ correlator is of the order of a few percent,
which is not distinguishable within the current level of uncertainties.

\subsection{Constraining the CME contribution}
\subsubsection{Describing the background with Blast-wave inspired LCC model}
As a first approach to constraining the CME contribution, a blast-wave
(BW) parametrisation~\cite{Retiere:2003kf} that describes the phase space
density at kinetic freeze-out, is used. This model assumes that the
radial expansion velocity is proportional to the distance from the
centre of the system and takes into account resonance production and
decays. Local charge conservation (LCC) is additionally incorporated
in this model by generating ensembles of particles with zero net
charge. The position of the sources of balancing charges are then
uniformly distributed within an ellipse. From now on this model will
be denoted as BW-LCC in the text. 

\begin{table}[h!]
\caption{List of the Blast-wave fit parameters.}
\begin{center}
\begin{tabular}{ |l|c|c|c|c| }
\hline
 Centrality & $T_{\rm kin}$ (MeV) & $\rho_{0}$ & $\rho_{2}$ &  $R_{x}/R_{y}$  \\ 
\hline
  0--5\%      & 91.3 $\pm$  3.5 & 1.26 $\pm$ 0.01 & 0.020 $\pm$ 0.001
                                                      & 0.956 $\pm$ 0.001 \\
\hline
5--10\%     & 87.0 $\pm$  3.5 & 1.27 $\pm$ 0.01 & 0.032 $\pm$ 0.001
                                                      &  0.933 $\pm$ 0.002 \\
\hline
10--20\%   & 84.8 $\pm$  4.9 & 1.25 $\pm$ 0.01 & 0.045 $\pm$ 0.003
                                                      & 0.905 $\pm$ 0.004 \\
\hline
20--30\%   & 87.4 $\pm$  4.8 & 1.23 $\pm$ 0.01 & 0.059 $\pm$ 0.007
                                                      & 0.872 $\pm$ 0.005 \\
\hline
30--40\%   & 91.6 $\pm$  3.8 & 1.20 $\pm$ 0.01 & 0.068 $\pm$ 0.003
                                                      & 0.844 $\pm$ 0.004 \\
\hline
40--50\%   & 95.1 $\pm$  3.3 & 1.15 $\pm$ 0.01 & 0.070 $\pm$ 0.003
                                                      & 0.823 $\pm$ 0.004 \\
\hline
50--60\%   & 98.1 $\pm$  3.2 & 1.09 $\pm$ 0.01 & 0.065 $\pm$ 0.002
                                                      & 0.807 $\pm$ 0.004 \\
\hline
60--70\%   & 108.0 $\pm$  3.2 & 0.99 $\pm$ 0.01 & 0.056 $\pm$ 0.002
                                                      & 0.786 $\pm$ 0.006 \\
\hline
\end{tabular}
\label{Tab:bwParameters}
\end{center}
\end{table}

\begin{figure}[t!h]
    \begin{center}
    \includegraphics[width=0.485\textwidth]{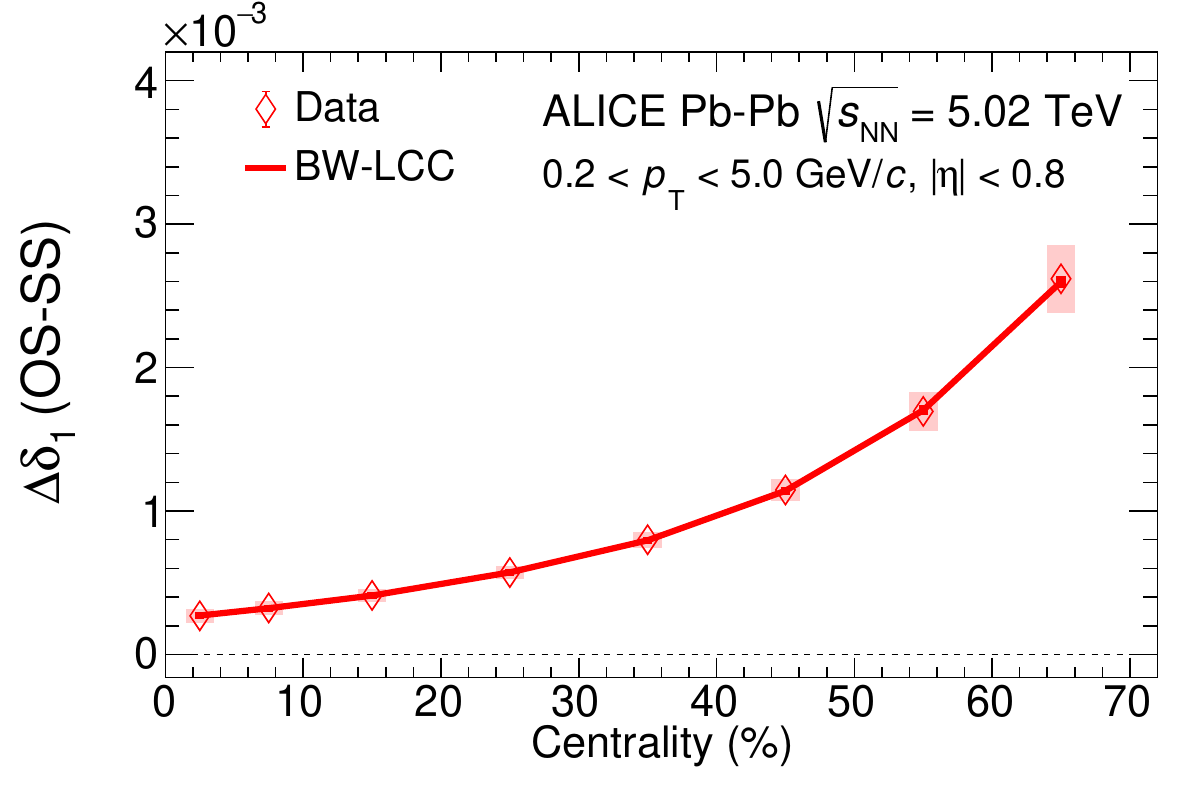}
    \includegraphics[width=0.485\textwidth]{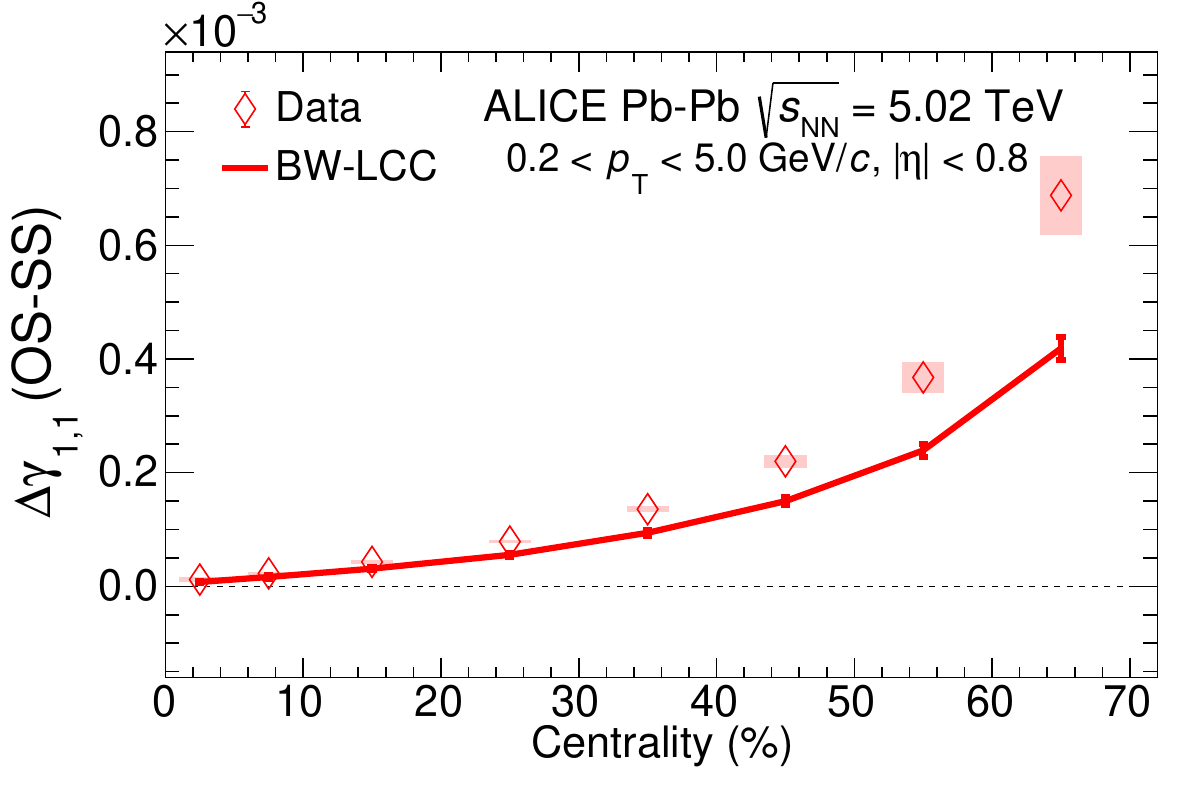}
    \end{center}
    \caption{(Left) The centrality dependence of $\Delta \delta_1$
      measured in Pb--Pb collisions at $\sqrt{s_\mathrm{{\rm NN}}} =
      5.02$~TeV. The curve (denoted as BW-LCC) presents the blast-wave
      parametrization coupled to local charge conservation effects.
      The model is tuned to reproduce the measured values of
      $\Delta \delta_1$ (see text for details). (Right) The comparison
      of the centrality dependence of the CME-sensitive correlator
      $\Delta \gamma_{1,1}$ with expectations from the BW-LCC model.}
 \label{fig:bw}
\end{figure}

Each particle of an ensemble is emitted by a fluid element with a
common collective velocity following the single-particle BW
parametrisation. The procedure starts from obtaining BW
parameters by fitting the $p_{\rm T}$ spectra~\cite{Acharya:2019yoi}
and the $p_{\rm  T}$-differential $v_{2}$ values~\cite{Acharya:2018zuq}
for charged pions, kaons, and protons (antiprotons) measured in Pb--Pb
collisions at $\sqrt{s_\mathrm{{\rm NN}}} = 5.02$~TeV. The fit ranges
for $p_{\rm T}$ spectra are 0.5 $\le p_{\rm T} \le$ 1.0 GeV/$c$, 0.2$\le
p_{\rm T} \le$ 1.5 GeV/$c$ and 0.2 $\le p_{\rm T} \le$ 3.0 GeV/$c$ for
pions, kaons and protons, respectively. The fit range for $p_{\rm
  T}$-differential $v_{2}$ and $v_{4}$ is 0.5 $\le p_{\rm T} \le$ 1.2
GeV/$c$.  
Table~\ref{Tab:bwParameters} presents the resulting BW
parameters, namely the kinetic freezeout temperature
($T_{\rm kin}$), radial flow ($\rho_{0}$) and its second order modulation 
($\rho_{2}$) as well as the spacial asymmetry ($R_{x}/R_{y}$). 
The next step required tuning the number of sources of balancing pairs
for each centrality interval to reproduce the centrality dependence of
$\Delta \delta_1$, the correlator which is mainly sensitive to
background effects. This procedure is repeated for every centrality
interval. The number of sources varies from $\sim$2476 to 193 for the
centrality intervals 0--5\% to 60--70\%.
The left panel of Fig.~\ref{fig:bw} presents the agreement achieved
between the measured results and the ones obtained from the
model. Overall the model describes the measurement fairly well with
deviations limited to $<$1\% for the whole centrality range. 
The tuned model is then used to extract the expectation for the
centrality dependence of the charge-dependent differences of the CME
sensitive correlator $\Delta \gamma_{1,1}$. The right panel of
Fig.~\ref{fig:bw} shows the comparison between the measured values of
$\Delta\gamma_{1,1}$ and estimates from the model. The estimate of
$\Delta\gamma_{1,1}$ from the model originates solely from the
contribution of local charge conservation effects coupled to elliptic
flow modulations. The curve underestimates the measured data points by
as much as $\approx$~39$\%$, with the disagreement increasing
progressively for more peripheral events.

\subsubsection{Describing the background with $v_{\mathrm n}$ and $\gamma_{\rm {m,n}}$}
\begin{figure}[t!h]
    \begin{center}
    \includegraphics[width=0.86\textwidth]{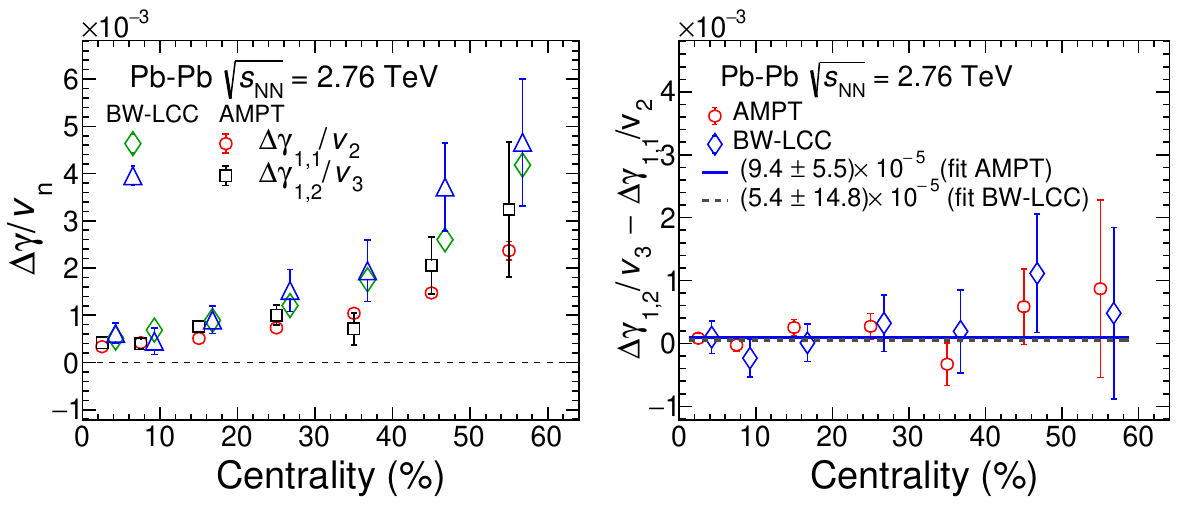}
    \end{center}
    \caption{(Left panel) The centrality dependence of $\Delta\gamma_{1,1}/v_2$ and 
    $\Delta\gamma_{1,2}/v_3$ for Pb--Pb collisions at $\sqrt{s_\mathrm{{\rm NN}}} = 2.76$~TeV 
    according to the AMPT and BW-LCC model. (Right panel) The
    differences between $\Delta\gamma_{1,1}/v_2$ and $\Delta\gamma_{1,2}/v_3$ 
    in AMPT and BW-LCC, denoted as $\Delta(\Delta\gamma/v_{\mathrm{n}})$.
    The solid and dotted line is the result of a fit with a constant
    function to AMPT and BW-LCC result respectively.}
 \label{fig:amptprediction}
\end{figure}

In the following, we attempt to constrain the background contribution
to the CME sensitive correlator $\gamma_{1,1}$ and thus give an
estimation of the fraction of the signal in Pb--Pb collisions. 
The approach described in Sec.~\ref{sec:factorization} relies on
the assumption that the coefficients $\kappa_{\mathrm{n}}$ have
similar magnitude, allowing one to calculate the background contribution
to $\Delta\gamma_{1,1}$, denoted as
$\Delta\gamma_{1,1}^{\textrm{Bkg}}$ from $\Delta\gamma_{1,2}$ 
according to Eq.~\ref{Eq:dataBkgC112}. This assumption was tested
using events produced with the string melting tune of A Multi Phase
Transport model (AMPT)~\cite{Zhang:1999bd,Lin:2000cx,Lin:2004en}. In
the string melting tune, the initial strings are melted into partons
whose interactions are described by a parton cascade
model~\cite{Zhang:1997ej}. These partons are then combined into  
final-state hadrons via a quark coalescence model. In this model, the
final-state hadronic rescattering is implemented including resonance
decays as well. The input parameters $\alpha_s = 0.33$ and a partonic
cross section of 1.5~mb were used to reproduce the centrality 
dependence of $v_2$ and $v_3$ for charged particles,
as reported in Ref.~\cite{Aamodt:2010pa}, for Pb--Pb events at
$\sqrt{s_\mathrm{{\rm NN}}} = 2.76$~TeV. About 40 million simulated 
Pb--Pb events were analysed, split into centrality based on the values
of the impact parameter. Only primary particles having the same kinematic 
selections as in the experimental data (i.e. $|\eta| < 0.8$ and
$0.2 < p_{\rm{T}} < 5$~GeV/$c$) were considered. The left panel of
Fig.~\ref{fig:amptprediction} presents the ratio of the
charge-dependent differences $\Delta\gamma_{1,1}$ and $\Delta\gamma_{1,2}$
to the relevant harmonics $v_2$ and $v_3$, respectively.
Similar ratios are also shown from the BW-LCC model, which is
discussed in the previous subsection.
The two sets of data points are compatible within uncertainties
over the entire centrality range for both AMPT and BW-LCC model.
This is also illustrated in the right panel of
Fig.~\ref{fig:amptprediction}, which presents the centrality
dependence of $(\Delta\gamma_{1,1}/v_2) - (\Delta\gamma_{1,2}/v_3)$,
denoted as $\Delta(\Delta\gamma/v_{\mathrm{n}})$.  
The corresponding data points from AMPT and BW-LCC are fitted with a
constant function, which yields a result compatible with zero within
the uncertainties of the fit i.e., $\Delta(\Delta\gamma/v_{\mathrm{n}})$ =
$(9.4 \pm 5.5) \times 10^{-5}$ and $(5.4 \pm 14.8) \times 10^{-5}$
respectively. 
This observation illustrates that within these models one can assume
$\kappa_2 \approx \kappa_3$. 
Results presented in this article have been also reproduced using the
AMPT version reported in~\cite{Choudhury:2019ctw}.    

The same procedure was used with Pb--Pb data recorded at both LHC
energies. Since the results of $\gamma_{2,2}$ do not give any
significant charge-dependent difference as a function of centrality
within statistical and systematic uncertainties (see Fig.~\ref{fig:Int3pcorr}),
only the values of $\gamma_{1,2}$ are used in the rest of the article
to estimate the background.

\begin{figure}[tb]
    \begin{center}
    \includegraphics[width =
    0.48\textwidth]{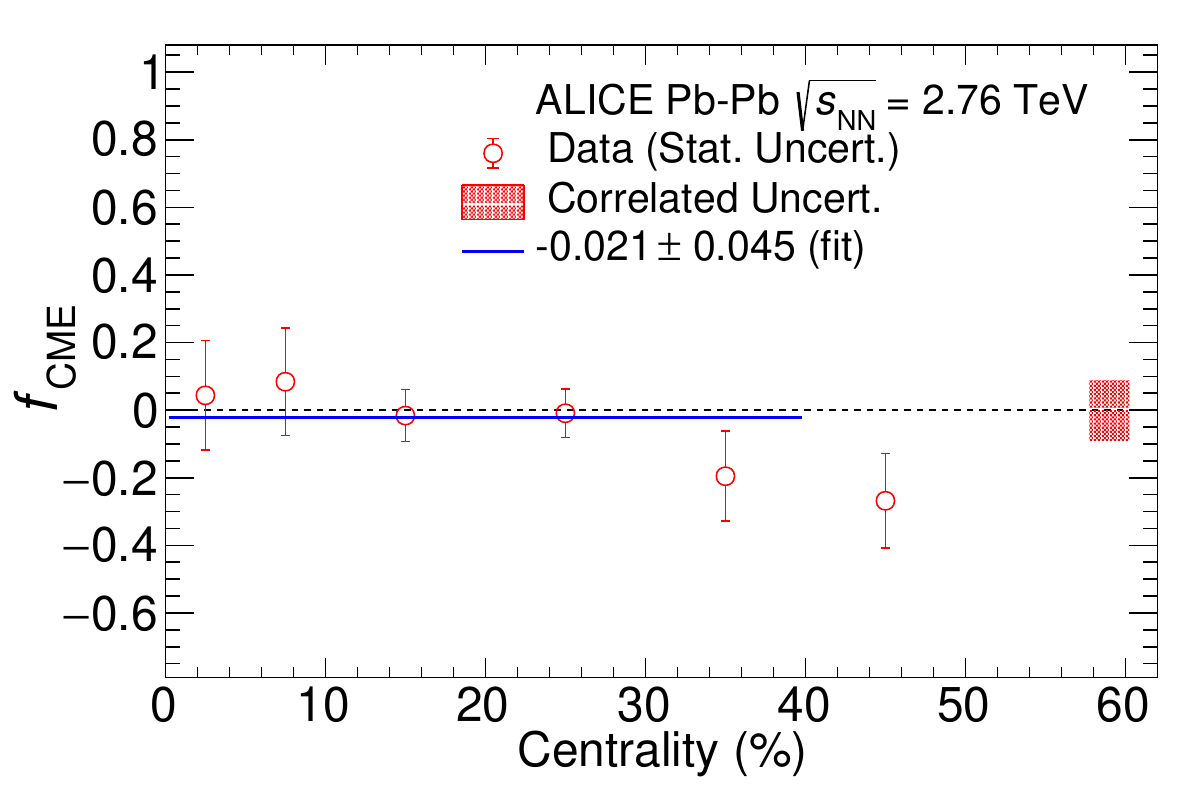}
    \includegraphics[width =
    0.48\textwidth]{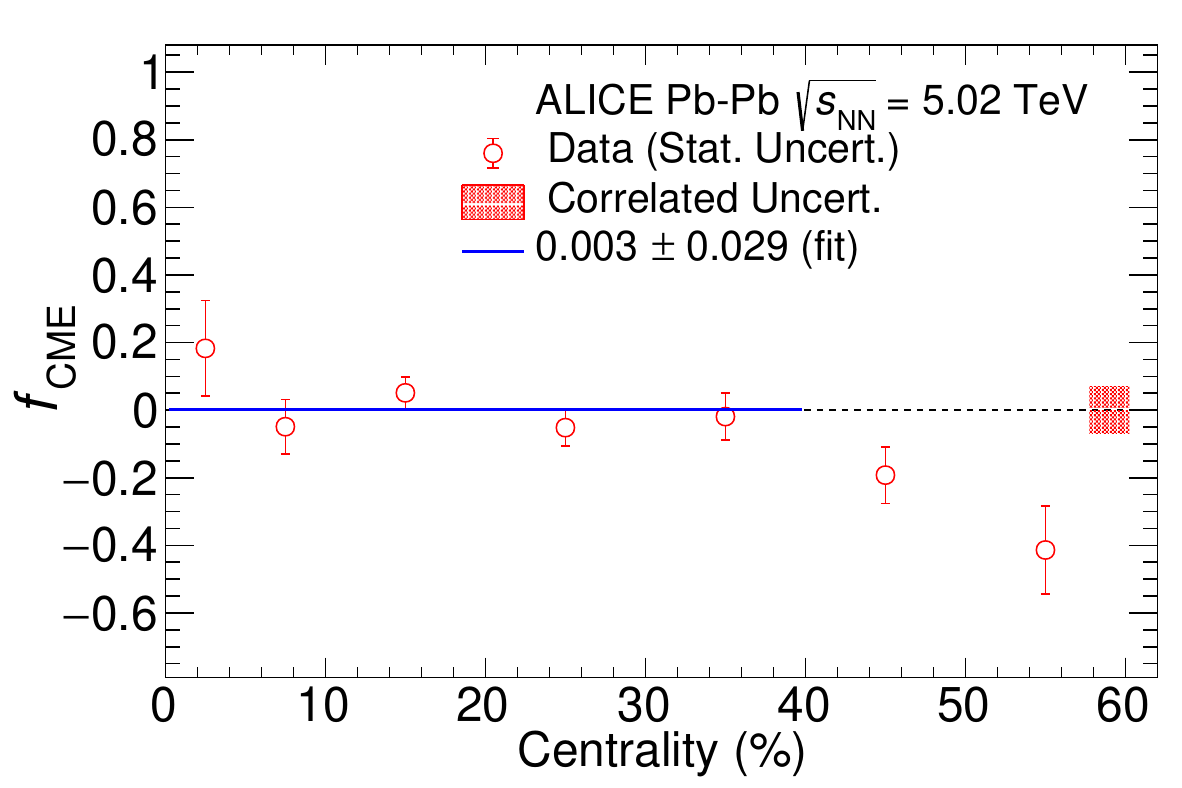}
    \end{center}
    \caption{(Left panel) The CME fraction extracted in Pb--Pb
      collisions at $\sqrt{s_\mathrm{{\rm NN}}} = 2.76$~TeV. 
     (Right panel) The CME fraction extracted in Pb--Pb collisions at
     $\sqrt{s_\mathrm{{\rm NN}}} = 5.02$~TeV. The systematic
     uncertainty is shown as hatched band at zero line around the centrality value of 60\%.
     The solid blue lines correspond to fit with a constant function
     to the data points. See text for details.}  
    \label{fig:cmeFraction}
\end{figure}

Although, CME sensitive $\gamma_{1,-3}$
correlator could have similar dependence on the background as $\gamma_{1,1}$
(i.e., $\gamma_{1,-3}\propto \delta_{1}v_{2}$), we observe in
Fig.~\ref{fig:Int3pcorr} that their magnitude are not the same. This
difference, which is being investigated as part a of future
publication, makes the $\gamma_{1,-3}$
correlator ambiguous to extract the CME contribution at this
stage. Therefore, in the following we only use $\gamma_{1,1}$
correlator to make quantitative measurement of CME.
The values of $v_2$ and  $v_3$ used to scale the charge-dependent
differences of $\gamma_{1,1}$ 
and $\gamma_{1,2}$, are the ones measured by ALICE in Pb--Pb
collisions reported in Ref.~\cite{Aamodt:2010pa} and
Ref.~\cite{Acharya:2018lmh} for $\sqrt{s_\mathrm{{\rm NN}}} =
2.76$~TeV and $\sqrt{s_\mathrm{{\rm NN}}} = 5.02$~TeV,
respectively. The value of $v_{2}$ is estimated as
the average of $v_{2}\{2\}$ and $v_{2}\{4\}$ to reduce
the biases due to fluctuations assuming a Gaussian probability
distribution~\cite{Voloshin:2007pc}.
Assuming $\kappa_2 \approx \kappa_3$ as supported by the model study,
the value of $\Delta \gamma_{1,2}\times v_2/v_3$ was used according to
Eq.~\ref{Eq:dataBkgC112} as a proxy for the magnitude of the
background contribution to the measurement of $\Delta \gamma_{1,1}$,
denoted as $\Delta \gamma_{1,1}^{\mathrm{Bkg}}$, for both LHC
energies. The CME fraction is then defined as 

\begin{equation}
f_{\rm CME} = 1 - \frac{\Delta \gamma_{1,1}^{\mathrm{Bkg} }}{\Delta \gamma_{1,1}},
\end{equation}

\noindent where $\Delta \gamma_{1,1}$ is the measured value of the
correlator presented on the left panel of Fig.~\ref{fig:DeltaGammaRun1Run2}. 
Figure~\ref{fig:cmeFraction} presents the centrality 
dependence of the CME fraction at $\sqrt{s_\mathrm{{\rm NN}}} = 2.76$~TeV
(left plot) and $\sqrt{s_\mathrm{{\rm NN}}} = 5.02$~TeV (right plot). The
systematic uncertainties on $f_{\rm CME}$ have been estimated from the
same sources as mentioned in Sec.~\ref{Section:Systematics}. In
addition, the contribution to the systematic uncertainty stemming from
$v_{2}$ was also estimated using $v_{2}\{2\}$ in
Eq.~\ref{Eq:dataBkgC112} instead of the average of $v_{2}\{2\}$ and
$v_{2}\{4\}$. All individual sources are then added in quadrature to
get the total systematic uncertainty. We have excluded any systematic
variation in the assumption of $\kappa_{2} \approx \kappa_{3}$ for the
reported values of $f_{\rm CME}$.  

The total, centrality-independent systematic uncertainty is indicated
by the shaded  box on the line at zero. It is seen that for both
energies $f_{\rm CME}$  is compatible with zero up to centrality
$\approx$ 40\%. For more peripheral collisions, the value of
$f_{\rm CME}$ is negative.
There could be several possible reasons behind this observation.
The event plane resolution starts to decline for peripheral events
giving rise to fluctuations in the measured observables. The non-flow, 
the effect of which is negligible for $\Delta\gamma_{\rm m,n}$ but
significant for $v_{\rm n}$, can affect the measurement for peripheral
events. There is another possibility that the ratio of the
proportionality constants, i.e, $\kappa_{2}/\kappa_{3}$ have a
centrality dependence instead of being a constant. Therefore the
observation of negative $f_{\rm CME}$ for peripheral events could be a
convoluted effect of all these underlying phenomena. 
However, within the scope of this
analysis, we have limited ourselves with constant behavior of
$\kappa_{2}/\kappa_{3}$ as a function of centrality.  


Finally, to estimate an upper limit on the contribution of the CME signal to the 
measurement of $\gamma_{1,1}$, the data points of $f_{\mathrm{CME}}$ are 
fitted with a constant function up to the 40\% centrality
interval. The fit yields values $-0.021 \pm 0.045$ and $0.003 \pm
0.029$ in Pb--Pb collisions at $\sqrt{s_{\rm NN}}$ = 2.76 and 5.02
TeV, respectively. These results are consistent with zero CME fraction
and correspond to upper limits on $f_{\mathrm{CME}}$ of 15--18\%
(20--24\%) at 95\% (99.7\%)  confidence level for the 0--40\%
centrality interval. The latter values are estimated assuming Gaussian
distributed uncertainties and taking into account that the CME
fraction has a lower bound of 0 (see Fig.~\ref{fig:cmeFraction}).  

\section{Summary} 
\label{Section:Summary}

In this article, we reported charge-dependent results
for various two-particle correlators as well as two-particle
correlations relative to different order symmetry planes. These
measurements are extracted from the analysis of Pb--Pb collisions
recorded by ALICE at $\sqrt{s_{\rm NN}}$ = 2.76 and 5.02 TeV. These
correlators exhibit different sensitivity to the signal induced by the
CME and to background effects, dominated by local charge conservations
coupled to anisotropic flow. 

All two-particle correlations of the form $\delta_{\mathrm{m}}$ for
$\mathrm{m}$~=~1, 2,~3~and~4 are dominated by background effects and
exhibit a significant centrality dependence. Among them, only $\delta_1$
exhibits a notable charge-dependent difference, which does  
not change significantly as a function of $\Delta p_{\mathrm{T}}$, but
increases with increasing $\overline{p_{\mathrm{T}}}$ and decreases
with increasing $\Delta \eta$ of the pair. For higher harmonics, on
the other hand, the charge-dependent differences become progressively
smaller and are compatible with zero for $\delta_3$ and $\delta_4$. 

The CME sensitive two-particle correlations relative to the second
order symmetry plane, $\gamma_{1,1}$ and $\gamma_{1,-3}$, exhibit a
significant charge-dependent difference, which increases towards peripheral
centrality intervals. Results on particle correlations relative to the
third order symmetry plane, expressed by $\gamma_{1,2}$, that probe
background effects associated with local charge conservation modulated
by triangular flow ($v_3$), also show a
significant charge-dependent difference, which increases for more
peripheral events. 
Finally, correlations between two particles relative to the fourth
order symmetry plane, that also have the potential to probe mainly
background effects, show no significant difference for pairs with
same and opposite electric charges, however they are suffering from
large statistical and systematic uncertainties.     

A blast wave parametrisation that incorporates local charge
conservation tuned to reproduce the components of the background, is
not able to fully describe the magnitude of the charge-dependent
differences of the CME-sensitive correlator $\gamma_{1,1}$. Finally,
the results of correlations relative to $\Psi_3$ and $\Psi_4$ that
probe mainly, if not solely, the contribution of the background
clearly show that these background effects are the dominating factor
to the measurements of $\gamma_{1,1}$.


\newenvironment{acknowledgement}{\relax}{\relax}
\begin{acknowledgement}
\section*{Acknowledgements}

The ALICE Collaboration would like to thank all its engineers and technicians for their invaluable contributions to the construction of the experiment and the CERN accelerator teams for the outstanding performance of the LHC complex.
The ALICE Collaboration gratefully acknowledges the resources and support provided by all Grid centres and the Worldwide LHC Computing Grid (WLCG) collaboration.
The ALICE Collaboration acknowledges the following funding agencies for their support in building and running the ALICE detector:
A. I. Alikhanyan National Science Laboratory (Yerevan Physics Institute) Foundation (ANSL), State Committee of Science and World Federation of Scientists (WFS), Armenia;
Austrian Academy of Sciences, Austrian Science Fund (FWF): [M 2467-N36] and Nationalstiftung f\"{u}r Forschung, Technologie und Entwicklung, Austria;
Ministry of Communications and High Technologies, National Nuclear Research Center, Azerbaijan;
Conselho Nacional de Desenvolvimento Cient\'{\i}fico e Tecnol\'{o}gico (CNPq), Financiadora de Estudos e Projetos (Finep), Funda\c{c}\~{a}o de Amparo \`{a} Pesquisa do Estado de S\~{a}o Paulo (FAPESP) and Universidade Federal do Rio Grande do Sul (UFRGS), Brazil;
Ministry of Education of China (MOEC) , Ministry of Science \& Technology of China (MSTC) and National Natural Science Foundation of China (NSFC), China;
Ministry of Science and Education and Croatian Science Foundation, Croatia;
Centro de Aplicaciones Tecnol\'{o}gicas y Desarrollo Nuclear (CEADEN), Cubaenerg\'{\i}a, Cuba;
Ministry of Education, Youth and Sports of the Czech Republic, Czech Republic;
The Danish Council for Independent Research | Natural Sciences, the VILLUM FONDEN and Danish National Research Foundation (DNRF), Denmark;
Helsinki Institute of Physics (HIP), Finland;
Commissariat \`{a} l'Energie Atomique (CEA) and Institut National de Physique Nucl\'{e}aire et de Physique des Particules (IN2P3) and Centre National de la Recherche Scientifique (CNRS), France;
Bundesministerium f\"{u}r Bildung und Forschung (BMBF) and GSI Helmholtzzentrum f\"{u}r Schwerionenforschung GmbH, Germany;
General Secretariat for Research and Technology, Ministry of Education, Research and Religions, Greece;
National Research, Development and Innovation Office, Hungary;
Department of Atomic Energy Government of India (DAE), Department of Science and Technology, Government of India (DST), University Grants Commission, Government of India (UGC) and Council of Scientific and Industrial Research (CSIR), India;
Indonesian Institute of Science, Indonesia;
Centro Fermi - Museo Storico della Fisica e Centro Studi e Ricerche Enrico Fermi and Istituto Nazionale di Fisica Nucleare (INFN), Italy;
Institute for Innovative Science and Technology , Nagasaki Institute of Applied Science (IIST), Japanese Ministry of Education, Culture, Sports, Science and Technology (MEXT) and Japan Society for the Promotion of Science (JSPS) KAKENHI, Japan;
Consejo Nacional de Ciencia (CONACYT) y Tecnolog\'{i}a, through Fondo de Cooperaci\'{o}n Internacional en Ciencia y Tecnolog\'{i}a (FONCICYT) and Direcci\'{o}n General de Asuntos del Personal Academico (DGAPA), Mexico;
Nederlandse Organisatie voor Wetenschappelijk Onderzoek (NWO), Netherlands;
The Research Council of Norway, Norway;
Commission on Science and Technology for Sustainable Development in the South (COMSATS), Pakistan;
Pontificia Universidad Cat\'{o}lica del Per\'{u}, Peru;
Ministry of Science and Higher Education, National Science Centre and WUT ID-UB, Poland;
Korea Institute of Science and Technology Information and National Research Foundation of Korea (NRF), Republic of Korea;
Ministry of Education and Scientific Research, Institute of Atomic Physics and Ministry of Research and Innovation and Institute of Atomic Physics, Romania;
Joint Institute for Nuclear Research (JINR), Ministry of Education and Science of the Russian Federation, National Research Centre Kurchatov Institute, Russian Science Foundation and Russian Foundation for Basic Research, Russia;
Ministry of Education, Science, Research and Sport of the Slovak Republic, Slovakia;
National Research Foundation of South Africa, South Africa;
Swedish Research Council (VR) and Knut \& Alice Wallenberg Foundation (KAW), Sweden;
European Organization for Nuclear Research, Switzerland;
Suranaree University of Technology (SUT), National Science and Technology Development Agency (NSDTA) and Office of the Higher Education Commission under NRU project of Thailand, Thailand;
Turkish Atomic Energy Agency (TAEK), Turkey;
National Academy of  Sciences of Ukraine, Ukraine;
Science and Technology Facilities Council (STFC), United Kingdom;
National Science Foundation of the United States of America (NSF) and United States Department of Energy, Office of Nuclear Physics (DOE NP), United States of America.
\end{acknowledgement}

\bibliographystyle{utphys}   
\bibliography{cmeMain}


\newpage
\appendix

%
%

\section{The ALICE Collaboration}
\label{app:collab}

\begingroup
\small
\begin{flushleft}
S.~Acharya\Irefn{org141}\And 
D.~Adamov\'{a}\Irefn{org95}\And 
A.~Adler\Irefn{org74}\And 
J.~Adolfsson\Irefn{org81}\And 
M.M.~Aggarwal\Irefn{org100}\And 
G.~Aglieri Rinella\Irefn{org34}\And 
M.~Agnello\Irefn{org30}\And 
N.~Agrawal\Irefn{org10}\textsuperscript{,}\Irefn{org54}\And 
Z.~Ahammed\Irefn{org141}\And 
S.~Ahmad\Irefn{org16}
% ...
\And
\renewcommand\labelenumi{\textsuperscript{\theenumi}~}

\section*{Affiliation notes}
\renewcommand\theenumi{\roman{enumi}}
\begin{Authlist}
\item \Adef{org*}Deceased
\item \Adef{orgI}Italian National Agency for New Technologies, Energy and Sustainable Economic Development (ENEA), Bologna, Italy
\item \Adef{orgII}Dipartimento DET del Politecnico di Torino, Turin, Italy
\item \Adef{orgIII}M.V. Lomonosov Moscow State University, D.V. Skobeltsyn Institute of Nuclear, Physics, Moscow, Russia
\item \Adef{orgIV}Department of Applied Physics, Aligarh Muslim University, Aligarh, India
\item \Adef{orgV}Institute of Theoretical Physics, University of Wroclaw, Poland
\end{Authlist}

\section*{Collaboration Institutes}
\renewcommand\theenumi{\arabic{enumi}~}
\begin{Authlist}
\item \Idef{org1}A.I. Alikhanyan National Science Laboratory (Yerevan Physics Institute) Foundation, Yerevan, Armenia
\item \Idef{org2}Bogolyubov Institute for Theoretical Physics, National Academy of Sciences of Ukraine, Kiev, Ukraine
\item \Idef{org3}Bose Institute, Department of Physics  and Centre for Astroparticle Physics and Space Science (CAPSS), Kolkata, India
\item \Idef{org4}Budker Institute for Nuclear Physics, Novosibirsk, Russia
\item \Idef{org5}California Polytechnic State University, San Luis Obispo, California, United States
\item \Idef{org6}Central China Normal University, Wuhan, China
\item \Idef{org7}Centre de Calcul de l'IN2P3, Villeurbanne, Lyon, France
\item \Idef{org8}Centro de Aplicaciones Tecnol\'{o}gicas y Desarrollo Nuclear (CEADEN), Havana, Cuba
\item \Idef{org9}Centro de Investigaci\'{o}n y de Estudios Avanzados (CINVESTAV), Mexico City and M\'{e}rida, Mexico
\item \Idef{org10}Centro Fermi - Museo Storico della Fisica e Centro Studi e Ricerche ``Enrico Fermi', Rome, Italy
\item \Idef{org11}Chicago State University, Chicago, Illinois, United States
\item \Idef{org12}China Institute of Atomic Energy, Beijing, China
\item \Idef{org13}Comenius University Bratislava, Faculty of Mathematics, Physics and Informatics, Bratislava, Slovakia
\item \Idef{org14}COMSATS University Islamabad, Islamabad, Pakistan
\item \Idef{org15}Creighton University, Omaha, Nebraska, United States
\item \Idef{org16}Department of Physics, Aligarh Muslim University, Aligarh, India
\item \Idef{org17}Department of Physics, Pusan National University, Pusan, Republic of Korea
\item \Idef{org18}Department of Physics, Sejong University, Seoul, Republic of Korea
\item \Idef{org19}Department of Physics, University of California, Berkeley, California, United States
\item \Idef{org20}Department of Physics, University of Oslo, Oslo, Norway
\item \Idef{org21}Department of Physics and Technology, University of Bergen, Bergen, Norway
\item \Idef{org22}Dipartimento di Fisica dell'Universit\`{a} 'La Sapienza' and Sezione INFN, Rome, Italy
\item \Idef{org23}Dipartimento di Fisica dell'Universit\`{a} and Sezione INFN, Cagliari, Italy
\item \Idef{org24}Dipartimento di Fisica dell'Universit\`{a} and Sezione INFN, Trieste, Italy
\item \Idef{org25}Dipartimento di Fisica dell'Universit\`{a} and Sezione INFN, Turin, Italy
\item \Idef{org26}Dipartimento di Fisica e Astronomia dell'Universit\`{a} and Sezione INFN, Bologna, Italy
\item \Idef{org27}Dipartimento di Fisica e Astronomia dell'Universit\`{a} and Sezione INFN, Catania, Italy
\item \Idef{org28}Dipartimento di Fisica e Astronomia dell'Universit\`{a} and Sezione INFN, Padova, Italy
\item \Idef{org29}Dipartimento di Fisica `E.R.~Caianiello' dell'Universit\`{a} and Gruppo Collegato INFN, Salerno, Italy
\item \Idef{org30}Dipartimento DISAT del Politecnico and Sezione INFN, Turin, Italy
\item \Idef{org31}Dipartimento di Scienze e Innovazione Tecnologica dell'Universit\`{a} del Piemonte Orientale and INFN Sezione di Torino, Alessandria, Italy
\item \Idef{org32}Dipartimento di Scienze MIFT, Universit\`{a} di Messina, Messina, Italy
\item \Idef{org33}Dipartimento Interateneo di Fisica `M.~Merlin' and Sezione INFN, Bari, Italy
\item \Idef{org34}European Organization for Nuclear Research (CERN), Geneva, Switzerland
\item \Idef{org35}Faculty of Electrical Engineering, Mechanical Engineering and Naval Architecture, University of Split, Split, Croatia
\item \Idef{org36}Faculty of Engineering and Science, Western Norway University of Applied Sciences, Bergen, Norway
\item \Idef{org37}Faculty of Nuclear Sciences and Physical Engineering, Czech Technical University in Prague, Prague, Czech Republic
\item \Idef{org38}Faculty of Science, P.J.~\v{S}af\'{a}rik University, Ko\v{s}ice, Slovakia
\item \Idef{org39}Frankfurt Institute for Advanced Studies, Johann Wolfgang Goethe-Universit\"{a}t Frankfurt, Frankfurt, Germany
\item \Idef{org40}Fudan University, Shanghai, China
\item \Idef{org41}Gangneung-Wonju National University, Gangneung, Republic of Korea
\item \Idef{org42}Gauhati University, Department of Physics, Guwahati, India
\item \Idef{org43}Helmholtz-Institut f\"{u}r Strahlen- und Kernphysik, Rheinische Friedrich-Wilhelms-Universit\"{a}t Bonn, Bonn, Germany
\item \Idef{org44}Helsinki Institute of Physics (HIP), Helsinki, Finland
\item \Idef{org45}High Energy Physics Group,  Universidad Aut\'{o}noma de Puebla, Puebla, Mexico
\item \Idef{org46}Hiroshima University, Hiroshima, Japan
\item \Idef{org47}Hochschule Worms, Zentrum  f\"{u}r Technologietransfer und Telekommunikation (ZTT), Worms, Germany
\item \Idef{org48}Horia Hulubei National Institute of Physics and Nuclear Engineering, Bucharest, Romania
\item \Idef{org49}Indian Institute of Technology Bombay (IIT), Mumbai, India
\item \Idef{org50}Indian Institute of Technology Indore, Indore, India
\item \Idef{org51}Indonesian Institute of Sciences, Jakarta, Indonesia
\item \Idef{org52}INFN, Laboratori Nazionali di Frascati, Frascati, Italy
\item \Idef{org53}INFN, Sezione di Bari, Bari, Italy
\item \Idef{org54}INFN, Sezione di Bologna, Bologna, Italy
\item \Idef{org55}INFN, Sezione di Cagliari, Cagliari, Italy
\item \Idef{org56}INFN, Sezione di Catania, Catania, Italy
\item \Idef{org57}INFN, Sezione di Padova, Padova, Italy
\item \Idef{org58}INFN, Sezione di Roma, Rome, Italy
\item \Idef{org59}INFN, Sezione di Torino, Turin, Italy
\item \Idef{org60}INFN, Sezione di Trieste, Trieste, Italy
\item \Idef{org61}Inha University, Incheon, Republic of Korea
\item \Idef{org62}Institute for Nuclear Research, Academy of Sciences, Moscow, Russia
\item \Idef{org63}Institute for Subatomic Physics, Utrecht University/Nikhef, Utrecht, Netherlands
\item \Idef{org64}Institute of Experimental Physics, Slovak Academy of Sciences, Ko\v{s}ice, Slovakia
\item \Idef{org65}Institute of Physics, Homi Bhabha National Institute, Bhubaneswar, India
\item \Idef{org66}Institute of Physics of the Czech Academy of Sciences, Prague, Czech Republic
\item \Idef{org67}Institute of Space Science (ISS), Bucharest, Romania
\item \Idef{org68}Institut f\"{u}r Kernphysik, Johann Wolfgang Goethe-Universit\"{a}t Frankfurt, Frankfurt, Germany
\item \Idef{org69}Instituto de Ciencias Nucleares, Universidad Nacional Aut\'{o}noma de M\'{e}xico, Mexico City, Mexico
\item \Idef{org70}Instituto de F\'{i}sica, Universidade Federal do Rio Grande do Sul (UFRGS), Porto Alegre, Brazil
\item \Idef{org71}Instituto de F\'{\i}sica, Universidad Nacional Aut\'{o}noma de M\'{e}xico, Mexico City, Mexico
\item \Idef{org72}iThemba LABS, National Research Foundation, Somerset West, South Africa
\item \Idef{org73}Jeonbuk National University, Jeonju, Republic of Korea
\item \Idef{org74}Johann-Wolfgang-Goethe Universit\"{a}t Frankfurt Institut f\"{u}r Informatik, Fachbereich Informatik und Mathematik, Frankfurt, Germany
\item \Idef{org75}Joint Institute for Nuclear Research (JINR), Dubna, Russia
\item \Idef{org76}Korea Institute of Science and Technology Information, Daejeon, Republic of Korea
\item \Idef{org77}KTO Karatay University, Konya, Turkey
\item \Idef{org78}Laboratoire de Physique des 2 Infinis, Ir\`{e}ne Joliot-Curie, Orsay, France
\item \Idef{org79}Laboratoire de Physique Subatomique et de Cosmologie, Universit\'{e} Grenoble-Alpes, CNRS-IN2P3, Grenoble, France
\item \Idef{org80}Lawrence Berkeley National Laboratory, Berkeley, California, United States
\item \Idef{org81}Lund University Department of Physics, Division of Particle Physics, Lund, Sweden
\item \Idef{org82}Nagasaki Institute of Applied Science, Nagasaki, Japan
\item \Idef{org83}Nara Women{'}s University (NWU), Nara, Japan
\item \Idef{org84}National and Kapodistrian University of Athens, School of Science, Department of Physics , Athens, Greece
\item \Idef{org85}National Centre for Nuclear Research, Warsaw, Poland
\item \Idef{org86}National Institute of Science Education and Research, Homi Bhabha National Institute, Jatni, India
\item \Idef{org87}National Nuclear Research Center, Baku, Azerbaijan
\item \Idef{org88}National Research Centre Kurchatov Institute, Moscow, Russia
\item \Idef{org89}Niels Bohr Institute, University of Copenhagen, Copenhagen, Denmark
\item \Idef{org90}Nikhef, National institute for subatomic physics, Amsterdam, Netherlands
\item \Idef{org91}NRC Kurchatov Institute IHEP, Protvino, Russia
\item \Idef{org92}NRC \guillemotleft Kurchatov\guillemotright~Institute - ITEP, Moscow, Russia
\item \Idef{org93}NRNU Moscow Engineering Physics Institute, Moscow, Russia
\item \Idef{org94}Nuclear Physics Group, STFC Daresbury Laboratory, Daresbury, United Kingdom
\item \Idef{org95}Nuclear Physics Institute of the Czech Academy of Sciences, \v{R}e\v{z} u Prahy, Czech Republic
\item \Idef{org96}Oak Ridge National Laboratory, Oak Ridge, Tennessee, United States
\item \Idef{org97}Ohio State University, Columbus, Ohio, United States
\item \Idef{org98}Petersburg Nuclear Physics Institute, Gatchina, Russia
\item \Idef{org99}Physics department, Faculty of science, University of Zagreb, Zagreb, Croatia
\item \Idef{org100}Physics Department, Panjab University, Chandigarh, India
\item \Idef{org101}Physics Department, University of Jammu, Jammu, India
\item \Idef{org102}Physics Department, University of Rajasthan, Jaipur, India
\item \Idef{org103}Physikalisches Institut, Eberhard-Karls-Universit\"{a}t T\"{u}bingen, T\"{u}bingen, Germany
\item \Idef{org104}Physikalisches Institut, Ruprecht-Karls-Universit\"{a}t Heidelberg, Heidelberg, Germany
\item \Idef{org105}Physik Department, Technische Universit\"{a}t M\"{u}nchen, Munich, Germany
\item \Idef{org106}Politecnico di Bari, Bari, Italy
\item \Idef{org107}Research Division and ExtreMe Matter Institute EMMI, GSI Helmholtzzentrum f\"ur Schwerionenforschung GmbH, Darmstadt, Germany
\item \Idef{org108}Rudjer Bo\v{s}kovi\'{c} Institute, Zagreb, Croatia
\item \Idef{org109}Russian Federal Nuclear Center (VNIIEF), Sarov, Russia
\item \Idef{org110}Saha Institute of Nuclear Physics, Homi Bhabha National Institute, Kolkata, India
\item \Idef{org111}School of Physics and Astronomy, University of Birmingham, Birmingham, United Kingdom
\item \Idef{org112}Secci\'{o}n F\'{\i}sica, Departamento de Ciencias, Pontificia Universidad Cat\'{o}lica del Per\'{u}, Lima, Peru
\item \Idef{org113}St. Petersburg State University, St. Petersburg, Russia
\item \Idef{org114}Stefan Meyer Institut f\"{u}r Subatomare Physik (SMI), Vienna, Austria
\item \Idef{org115}SUBATECH, IMT Atlantique, Universit\'{e} de Nantes, CNRS-IN2P3, Nantes, France
\item \Idef{org116}Suranaree University of Technology, Nakhon Ratchasima, Thailand
\item \Idef{org117}Technical University of Ko\v{s}ice, Ko\v{s}ice, Slovakia
\item \Idef{org118}The Henryk Niewodniczanski Institute of Nuclear Physics, Polish Academy of Sciences, Cracow, Poland
\item \Idef{org119}The University of Texas at Austin, Austin, Texas, United States
\item \Idef{org120}Universidad Aut\'{o}noma de Sinaloa, Culiac\'{a}n, Mexico
\item \Idef{org121}Universidade de S\~{a}o Paulo (USP), S\~{a}o Paulo, Brazil
\item \Idef{org122}Universidade Estadual de Campinas (UNICAMP), Campinas, Brazil
\item \Idef{org123}Universidade Federal do ABC, Santo Andre, Brazil
\item \Idef{org124}University of Cape Town, Cape Town, South Africa
\item \Idef{org125}University of Houston, Houston, Texas, United States
\item \Idef{org126}University of Jyv\"{a}skyl\"{a}, Jyv\"{a}skyl\"{a}, Finland
\item \Idef{org127}University of Liverpool, Liverpool, United Kingdom
\item \Idef{org128}University of Science and Technology of China, Hefei, China
\item \Idef{org129}University of South-Eastern Norway, Tonsberg, Norway
\item \Idef{org130}University of Tennessee, Knoxville, Tennessee, United States
\item \Idef{org131}University of the Witwatersrand, Johannesburg, South Africa
\item \Idef{org132}University of Tokyo, Tokyo, Japan
\item \Idef{org133}University of Tsukuba, Tsukuba, Japan
\item \Idef{org134}Universit\'{e} Clermont Auvergne, CNRS/IN2P3, LPC, Clermont-Ferrand, France
\item \Idef{org135}Universit\'{e} de Lyon, Universit\'{e} Lyon 1, CNRS/IN2P3, IPN-Lyon, Villeurbanne, Lyon, France
\item \Idef{org136}Universit\'{e} de Strasbourg, CNRS, IPHC UMR 7178, F-67000 Strasbourg, France, Strasbourg, France
\item \Idef{org137}Universit\'{e} Paris-Saclay Centre d'Etudes de Saclay (CEA), IRFU, D\'{e}partment de Physique Nucl\'{e}aire (DPhN), Saclay, France
\item \Idef{org138}Universit\`{a} degli Studi di Foggia, Foggia, Italy
\item \Idef{org139}Universit\`{a} degli Studi di Pavia, Pavia, Italy
\item \Idef{org140}Universit\`{a} di Brescia, Brescia, Italy
\item \Idef{org141}Variable Energy Cyclotron Centre, Homi Bhabha National Institute, Kolkata, India
\item \Idef{org142}Warsaw University of Technology, Warsaw, Poland
\item \Idef{org143}Wayne State University, Detroit, Michigan, United States
\item \Idef{org144}Westf\"{a}lische Wilhelms-Universit\"{a}t M\"{u}nster, Institut f\"{u}r Kernphysik, M\"{u}nster, Germany
\item \Idef{org145}Wigner Research Centre for Physics, Budapest, Hungary
\item \Idef{org146}Yale University, New Haven, Connecticut, United States
\item \Idef{org147}Yonsei University, Seoul, Republic of Korea
\end{Authlist}
\endgroup
  
\end{document}